\documentclass[%
 aip,
 amsmath,amssymb,
reprint,
]{revtex4-1}

\usepackage{graphicx}
\usepackage{amsmath}
\usepackage{etoolbox}
\usepackage[version=3]{mhchem} 
\usepackage{wasysym}
\usepackage{amsmath,amssymb,latexsym}
\usepackage{pifont}
\usepackage[dvipsnames]{xcolor}
\usepackage{bm}

\usepackage{algorithm}
\usepackage{algpseudocode}
\usepackage{dsfont}

\usepackage{bbold}

\begin{document}

\title{Chemical Potential of a Flexible Polymer Liquid in a Coarse-Grained Representation}

\author{M. Dinpajooh}
\affiliation{Department of Chemistry and Biochemistry, University of Oregon, Eugene, Oregon 97403, USA}
\affiliation{Present address: Physical and Computational Sciences Directorate, Pacific Northwest National Laboratory, WA, 99352 USA}
\author{J. Millis}
\affiliation{Department of Physics, University of Oregon, Eugene, Oregon 97403, USA}
\author{J. P. Donley}
\affiliation{Material Science Institute, University of Oregon, Eugene, Oregon 97403, USA}
\author{M. G. Guenza}
\email{mguenza@uoregon.edu}
\affiliation{Department of Chemistry and Biochemistry, University of Oregon, Eugene, Oregon 97403, USA}
\affiliation{Department of Physics, University of Oregon, Eugene, Oregon 97403, USA}
\affiliation{Material Science Institute, University of Oregon, Eugene, Oregon 97403, USA}
\altaffiliation{Institute for Fundamental Science, University of Oregon, Eugene, Oregon 97403, USA}

\date{\today}

\begin{abstract}
While the excess chemical potential is the key quantity in determining phase diagrams, its direct computation for high-density liquids of long polymer chains has posed a significant challenge. Computationally, the excess chemical potential is calculated using the Widom insertion method, which involves monitoring the change in internal energy as one incrementally introduces individual molecules in the liquid. However, when dealing with dense polymer liquids, inserting long chains requires generating trial configurations with a bias that favors those at low energy on a unit-by-unit basis: a procedure that becomes more challenging as the number of units increases. Thus, calculating the excess chemical potential of dense polymer liquids using this method becomes computationally intractable as the chain length exceeds $N \geq 30$.
Here, we adopt a coarse-grained model derived from integral equation theory, for which inserting long polymer chains becomes feasible. The Integral Equation theory of Coarse-Graining (IECG) represents a polymer as a sphere or a collection of blobs interacting through a soft potential. 
We employ the IECG approach to compute the excess chemical potential using Widom's method for polymer chains of increasing lengths, extending up to $N=720$ monomers, and at densities reaching up to $\rho= 0.767$ g/cm$^{3}$.
From a fundamental perspective, we demonstrate that the excess chemical potentials remain nearly constant across various levels of coarse-graining, offering valuable insights into the consistency of this type of procedure.
Ultimately, we argue that current Monte Carlo (MC) algorithms, originally designed for atomistic simulations, such as Configurational Bias Monte Carlo (CBMC) methods, can significantly benefit from the integration of the IECG approach, thereby enhancing their performance in the study of phase diagrams of polymer liquids.
\end{abstract}

\maketitle

\section{Introduction}

To gain a comprehensive understanding of the thermodynamic stability of polymer melts, it is crucial to precisely determine their phase diagram by calculating their chemical potential.\cite{Zoback1993} Nevertheless, conducting an experimental investigation to determine the phase diagram of polymer chain melts is often unattainable due to the critical temperature of the polymer melt typically exceeding its degradation temperature. In essence, the polymer degrades before its coexistence curve can be measured. In response, computational modeling has emerged as a solution to determine polymer coexistence curves, effectively circumventing the challenges posed by experimental phase diagram determination.\cite{Frenkel2002}  The isothermal coexistence of two phases is defined by equal pressure and equal chemical potential in the two phases ($P_{\rho(1)}=P_{\rho(2)}$ and $\mu_{\rho(1)}=\mu_{\rho(2)}$), making accurate numerical calculations of pressure and chemical potential indispensable. Nonetheless, this task can prove to be non-trivial.

Calculating pressure from the simulation trajectory using the virial equation is straightforward, but determining the chemical potential can be challenging. The chemical potential, being a free energy, is calculated as an excess quantity by taking the difference from a known reference state, typically the gas phase. While determining the free energy of a system in its gas phase is not problematic, the challenging part lies is calculating the contribution of the excess chemical potential to the free energy.

The theoretical framework for calculating the excess chemical potential as an energy difference was first developed by Ben Widom. In Widom's method, the chemical potential is computed by measuring the difference in internal energy when inserting a 'ghost' molecule into the gas or liquid matrix.\cite{Wldom1963,Frenkel2002,Widom1978} This method is computationally straightforward and easily implemented in Monte Carlo (MC) simulations of liquids. However, the one-step insertion of the molecule becomes impractical when inserting a sizable molecule, such as a polymer, into a dense liquid, as it is unlikely to be accepted.

Throughout the years, various sophisticated computational techniques have been developed to enhance the insertion probability of polymer chains into a dense matrix. Among these techniques, the configurational-bias Monte Carlo (CBMC) method by Siepmann and coworkers has been the most effective.\cite{Siepmann1990} In the CBMC method, a polymer chain is gradually inserted into the polymer matrix, one monomer at a time.  After placing the first monomer inside the matrix at a random location, the second monomer is inserted by seeking an unoccupied site within a spatial region surrounding the first monomer.  This process is repeated until the entire chain is inserted. The CBMC method allows the direct calculation of the excess chemical potential from the chain insertion probability, following Rosenbluth statistics.\cite{Rosenbluth1955} Originally developed for a polymer matrix on a lattice, the chain insertion method has been extended to a continuum by de Pablo and coworkers,\cite{DePablo1992} and to extensible chains by Mooij and Frenkel.\cite{Mooij1994}  In contrast to Widom's simple direct insertion of the entire molecule in one step,\cite{Widom1978} the CBMC method offers a higher probability of successfully inserting the polymer chain. However,  as the chain length increases, finding an available site becomes progressively more challenging.\cite{Vlugt1998} Consequently, polymers with more than $30$ monomers can hardly be inserted into dense polymer melts, leading to limitations in determining the chemical potential for long polymer chains. For instance, Siepmann and colleagues investigated the vapor liquid coexistence curves in samples featuring degrees of polymerization up to $N=16$ for both \textit{n}-alkanes and perfluorinated alkanes.\cite{Martin1998c,Cui1998} More recently, their research extended to examining head-to-head polypropylene up to $N=36$ monomers, as well as polypropylene with the same monomer count, alongside poly(ethylene-\textit{alt}-propylene) consisting of $N=30$ monomers.\cite{Chen2016}

To enhance the capabilities of MC configurational sampling by chain insertion, several alternative methods have been proposed, including the pruned-enriched Rosenbluth method,\cite{Grassberger1997a} and the incremental gauge cell method.\cite{Rasmussen2011} Building upon the CBMC method, Kumar and coworkers focused on solving the incremental chemical potential, which represents the increase in excess chemical potential with the addition of a single monomer segment. The \textit{intramolecular} chemical potential of the full chain is thus calculated segment by segment. Notably, this potential becomes independent of chain length for chains longer than ten segments.  Note that the incremental chemical potential was reported for dilute solutions of chains with up to $30$ monomers, thus for lower densities and shorter chains than those considered in our study.\cite{Kumar1991,Kumar1992a}  

An expanded ensemble method was proposed by Wilding and M\"{u}ller, in which a chain is introduced as a ``permanent penetrable chain" that is gradually coupled to or decoupled from the matrix by tuning an interaction parameter.\cite{Wilding1994} In their 'expanded ensemble simulation' polymers were modeled on a lattice using the bond fluctuation model. For an athermal polymer system, they simulated chains with length up to $N=80$ at a volume fraction of $0.4$, significantly lower than the melt density considered in this study. Additionally, they conducted simulations of thermal polymer chains with degrees of polymerization up to $N=40$ at a volume fraction of $0.2$. Later, Escobedo and de Pablo introduced the EVALENCH method, which combines elements of M\"{u}ller's expanded ensemble, Kumar's chain increment, and Siepmann's continuum configurational-bias method. This approach was applied to calculate the chemical potential for tangent hard spheres with $4$, $8$, $16$, and $32$ monomers at low densities, and the results aligned with those previously reported by other methods.\cite{Escobedo1995} 

A number of other sophisticated schemes were proposed by Theodorou and coworkers to extend particle deletion methods,\cite{Spyriouni1997,Spyriouni1998,Galata2018} where the excess chemical potential was decomposed into an energetic contribution and a volume term related to cavity formation. In the staged particle deletion method, the excess chemical potential was calculated through a combination of insertion and deletion moves, chain breaking and rebuilding, and more. Using the particle deletion method together with the connectivity-altering end-bridging Monte Carlo move to make an equilibrated matrix, Theodorou and coworkers investigated the infinite dilution solubility of benzene in a matrix of polyethylene chains with degree of polymerization $N=44$.\cite{DeAngelis} The phase diagram of long chains with $N=2048$ monomers in \textit{semi-dilute solution} was calculated with the pruned-enriched Rosenbluth method for polymer chains modeled as self-avoiding walks with attractive energy between each pair of neighboring nonbonded monomers on a simple cubic lattice
with helical boundary conditions.\cite{diluteplo} 

A common observation from these studies is that they involve polymeric systems at low density, or in solution.\cite{diluteplo} Computational investigations involving realistic polymer structures with all-atom or United Atom (UA) resolutions at high density remain computationally intractable when the chain length exceeds $30$ monomers.\cite{Chen2016} This absence of computational research addressing the phase diagram of high-density liquids containing long polymer chains underscores the challenges associated with measuring converging excess free energies for these systems using insertion methods. 

An alternative computational method to Widom insertion is thermodynamic integration, where the free energy is calculated by integrating the pressure along a trajectory of increasing density.\cite{Frenkel2002} Nevertheless, this approach remains computationally demanding for polymeric liquids, because it is sensitive to the chosen integration path, and its accuracy hinges on factors such as sampling frequency, integration point placement, and the precision of the initial integration state. In the presence of a phase transition, thermodynamic integration becomes impractical due to the challenge of defining a reliable integration path that captures necessary ensemble averages while transitioning between phases.\cite{Frenkel2002}

Some methods rely on determining the equation of state (EOS)
from which the chemical potential can be calculated using standard thermodynamic relations.\cite{Prausnitz-1986}
For instance, methods have been developed using a perturbative expansion around the hard-sphere EOS. However, such a procedure becomes challenging for complex systems because it requires a significant number of adjustable parameters to fit for accuracy.\cite{Gross2002,DeAngelis} 
Thus, the Widom method could still be the most promising strategy to solve this problem.

Another theoretical method is self-consistent field theory. This is a continuum approach that effectively predicts phase diagrams of complex polymeric systems in the limit of high molecular weight chains ($N \rightarrow \infty$) and high-density liquids. In this context, fluctuations are suppressed, and the specific monomeric structure of the sample becomes less significant. Field-theoretical calculations rely on phenomenological interaction parameters and do not directly connect to the atomistic picture.  Compared with mean-field and continuum theories, particle-based methods hold an advantage by directly connecting to the local structural information while incorporating fluctuations.\cite{Baeurle2006}

Our proposed approach combines Widom's one-step, one-molecule insertion technique with a coarse-grained (CG) representation of polymer liquids based on the Integral Equation theory of Coarse-Graining (IECG).\cite{Guenza2018} The IECG method is an extension of polymer Reference Interaction Site Model (polymer RISM or PRISM) theory, which is itself an extension of atomic Ornstein-Zernike (OZ)  integral equation theory.\cite{schweizer_prism_1994} Notably, the soft potential inherent to the IECG method addresses the challenges associated with insertion, particularly for longer polymer chains where the interpolymer potential of IECG becomes softer.  This results in the efficient acceptance of most chain insertions during Monte Carlo simulations, facilitating the straightforward computation of the excess chemical potential.

To calculate the excess chemical potential using the IECG approach, we integrated our CG potential into an open source software package called Monte Carlo for Complex Chemical Systems (MCCCS),\cite{Siepmann1990,Martin2013} created and maintained by the Siepmann group, that implements a CBMC algorithm within the Gibbs ensemble Monte Carlo (GEMC) simulation method.\cite{Panagiotopoulos1987a} This modified code, referred to herein as ``MCCCS-CG", allowed direct calculation of the chemical potential for polymer liquids at atomistic resolutions using the CBMC approach and was then extended to evaluate the excess chemical potential of CG polymer chains across various state points for a range of polymer liquids at high densities. These thermodynamic conditions ensured the stability of the liquid phase without spontaneous phase transitions.

To validate the accuracy of our proposed method, we undertook two distinct procedures. Initially, we assessed the robustness of the IECG description within MC simulations by comparing the structural and thermodynamic attributes of equilibrated polymer liquids with corresponding atomistic Molecular Dynamics (MD) simulations under the same thermodynamic conditions. This step established the faithful reproduction of structural features, such as pair distribution functions, and thermodynamic aspects like pressure, confirming the fidelity of the MCCCS-CG code in emulating the properties of atomistic MD simulations.

As a secondary test, we evaluated the MCCCS-CG simulation data for the excess chemical potential by comparing them to theoretical predictions for the same quantity. The IECG method provided a theoretical expression amenable to numerical solution across any thermodynamic state point, and an analytical solution valid, at the very least, in the high-density liquid limit. In this way, we confirmed the consistency of the IECG approach in calculating the excess chemical potential for high-density polymer liquids. The success of these tests indicates that the proposed method could be employed to predict phase diagrams of polymer melts with long chains because it provides a reliable evaluation of the excess chemical potential in the high-density regime, where insertion methods using atomistic polymer resolution do not perform well. 

Importantly, we examined the consistency of the chemical potential for various levels of CG, meaning the representation of polymer chains as different numbers of CG units, denoted as $n_b$. We demonstrated that the excess chemical potential is invariant for practical values of $n_b$.
We stress though that while both the IECG theory and simulation provide values for the excess chemical potential, only the CG simulation is able to provide the molecular details necessary for obtaining atomistic configuration data through backmapping, if desired. Lastly, we emphasize the significant computational efficiency of our approach, which is orders of magnitude faster than atomistic simulations. In this study, we computed the excess chemical potential for high-density polyethylene melts encompassing degrees of polymerization ranging from $N=44$ to $N=720$. The agreement between theoretical predictions and simulations consistently proved to be satisfactory across all sample cases.

The remainder of this manuscript is structured as follows. We first describe our model system and derive an exact statistical mechanical expression for the chemical potential following the Widom approach. Then, we discuss the details of the intramolecular and intermolecular energies used in the expression for the chemical potential, including their connection to the IECG theory.  We then summarize the IECG method of determining elements in these energies by fitting a few parameters to short-time atomistic simulation data. Following this, we derive simple forms for the excess chemical potential that depend only on functions already established in the IECG theory. Subsequently, we discuss the ingredients of the associated CG simulation and how it extends prior simulation codes. We then present and compare the results obtained from the IECG theory and simulation. Lastly, we provide a summary of the work and discuss limitations and advatages of the IECG approach.

\section{Theory}
\subsection{Model system and formal expressions for the chemical potential}
\label{theory}

Consider a liquid of $n$ polymers in a volume $V$ and at a temperature $T$. Each polymer has a degree of polymerization $N$, partitioned into a variable number of CG units or blobs, $n_b$, where each unit contains a number $N_b=N/n_b$ of monomers.
The chain number density is then $\rho_{ch}=n/V$, the monomer density is $\rho=\rho_{ch}N$, and the density of CG units is $\rho_b=\rho/N_b$.

To calculate the chemical potential we follow Widom's insertion method, where we measure the change in Helmholtz free energy ($F$) when  we ``virtually" add one polymer chain, identical to the other chains, to the liquid in the canonical ensemble (constant $n$, $V$ and $T$). When this ghost polymer is inserted into a matrix of $n-1$ host polymers, the change in free energy is the blob-blob chemical potential ($\mu^{bb}$) 
\begin{eqnarray}
\label{mu0}
\mu^{bb}&=&\left( \frac{\partial F^{bb}}{\partial n}   \right)_{V,T}=-k_{\rm B}T \ln{\frac{Q_{nn_b}}{Q_{(n-1)n_b}}} \ ,
\end{eqnarray}
where $k_{\rm B}$ is Boltzmann's constant, and  $Q_{nn_b}$ denotes the canonical partition function of a liquid of $n$ CG chains, each containing $n_b$ blobs,
\begin{equation}
\label{partf}
Q_{n n_b}=\frac{Z^{bb}_{n}}{\Lambda^{3n n_b} n!} \  .
\end{equation}
Here, $Z^{bb}_n$ is the blob-blob configurational integral, and $\Lambda=( 2 \pi \beta \hbar^2/m_b)^{1/2}$ is the mean thermal de Broglie wavelength of the blob, where $m_b$ is the mass of a blob, $\hbar$ is Planck's constant divided by $2\pi$, and $\beta = 1/(k_{\rm B}T)$.

The configurational integral in turn is
\begin{equation}
\label{Zbb}
Z^{bb}_{n}=\int {\cal D}\mathbf{R}_n  \exp\left [{-\beta (U^{bb}_{\rm intra}(\mathbf{R}_n) + U^{bb}_ {\rm inter}(\mathbf{R}_n))}\right ] \ ,
\end{equation}
where $\mathbf{R}_n \equiv \{\mathbf{r}_{11},\cdots ,\mathbf{r}_{1n_b},\cdots ,\mathbf{r}_{n1},\cdots,\mathbf{r}_{nn_b}\}$ denotes a configuration of the CG site positions of the $n$ chains, with $\mathbf{r}_{i\alpha}$ the position of site $\alpha$ on polymer $i$.  As such, $\int {\cal D}\mathbf{R}_n \equiv \int \prod_{i=1}^n\prod_{\alpha=1}^{n_b} d\mathbf{r}_{i\alpha}$ denotes an integration over the entire space of site positions.  $U^{bb}_{\rm intra}(\mathbf{R}_n)$ and $U^{bb}_{\rm inter}(\mathbf{R}_n)$ are the total intramolecular and the intermolecular internal energies of the $n$-chain liquid, respectively, in the configuration $\mathbf{R}_n$. 

Incorporating Eqs.\eqref{partf} and \eqref{Zbb} into Eq.\eqref{mu0} yields  
\begin{equation}
\label{mu_tot}
\mu^{bb} = \mu^{bb}_{\rm id} + \mu^{bb}_{\rm exc}.
\end{equation}
Here, $\mu^{bb}_{\rm id}$ is the chemical potential needed to insert one chain into an ideal gas of identical chains, and is given by
\begin{equation}
\label{mu_id}
\mu^{bb}_{\rm id} = - k_{\rm B} T \ln{\frac{Z^{bb}_{1}}{\Lambda^{3 n_b} n}},
\end{equation}
while  $\mu^{bb}_{\rm exc}$, the excess chemical potential, accounts for the intermolecular interactions between the single inserted chain and the other polymer chains in the liquid, and the change of conformation of the single chain when it is inserted into the liquid, given by
\begin{equation}
\label{mu_exc}
\mu^{bb}_{\rm exc} =- k_{\rm B} T \ln{\frac{Z^{bb}_{n}}{Z^{bb}_{(n-1)}Z^{bb}_{1}}} .
\end{equation}
 Note that  $Z^{bb}_1$ is the configurational integral over the coordinates of a single isolated polymer chain  in the volume $V$. An approximation to Eq.\eqref{mu_exc} will be derived below.

\subsection{Intramolecular and Intermolecular internal energies in the IECG theory}
\label{IECGformalism}

To determine the partition function of a polymer liquid in a CG representation, $Z^{bb}_n$, given by Eq.\eqref{Zbb}, expressions for the intramolecular and intermolecular energies, $U_{\rm intra}^{bb}(\mathbf{R}_n)$ and $U_{\rm inter}^{bb}(\mathbf{R}_n)$, respectively, are needed. To that end, we utilize the integral equation theory of CG (IECG). The IECG formalism is based upon the polymer reference interaction site model (polymer-RISM or PRISM) theory\cite{Schweizer1997}, which itself is a generalization of atomic Ornstein-Zernike (OZ) theory\cite{j_p_hansen_theory_2013} to macromolecules. 

In the IECG theory, the chain is typically modeled with only a few blobs, so that for long flexible chains with $N_b \gg 1$, the correlations between blobs can be modeled as Gaussian. The average square end-to-end distance at melt densities is then  $\langle R^2\rangle =(n_b-1)l^2$, where $l$ is the average distance (bond length) between CG units.

If long-range interactions within a chain can be ignored, say, due to inter-chain screening at melt densities, then at the simplest level the chain structure is ideal. In this case, the intramolecular potential, $U_{\rm intra}^{bb}(\mathbf{R}_n)$, can be represented as a sum of Edwards Hamiltonians\cite{Doi1988}
\begin{equation}
\label{edwardsH}
U_{\rm Ed}^{bb}(\mathbf{R}_n) = k_{\rm B}T\sum_{i=1}^n\sum_{\alpha=1}^{n_b-1} \frac{3}{2} (l_{i\alpha}/l)^2,
\end{equation}
where $l_{i\alpha} \equiv \vert \mathbf{r}_{i,\alpha} - \mathbf{r}_{i,\alpha+1}\vert$ is the distance between neighboring CG sites, $\alpha$ and $\alpha+1$, on chain $i$.
Using this in Eq.\eqref{Zbb}, the single molecule partition function, $Z^{bb}_1$ can be calculated analytically to be
\begin{equation}
\label{Zbb1}
Z^{bb}_1 = V \left (2\pi l^2/3 \right )^{3/2(n_b-1)}.
\end{equation}
This expression depends on temperature only weakly through the bond length $l$, which also depends upon $n_b$ since $\langle R^2\rangle$ is invariant.

In general though, long-range interactions cannot be ignored as they change the chain structure from ideal to self-avoiding at low (or zero) density, which is the regime in which $Z^{bb}_1$ is computed. The IECG theory then uses a more sophisticated form for $U_{\rm intra}^{bb}(\mathbf{R}_n)$, which we state below. For a detailed description of these expressions, and also those for $U_{\rm inter}^{bb}(\mathbf{R}_n)$, please refer to our previously published papers.\cite{Dinpajooh2018,Dinpajooh2019,Clark2015,McCarty2014,Clark2012}

The IECG form for $U_{\rm intra}^{bb}(\mathbf{R}_n)$ contains contributions from bond, angle, and non-bonded long-range interactions. The dihedral contribution is not relevant in this case because the CG sites are soft and can overlap. So,
\begin{equation}
\label{Uintra}
U^{bb}_{\rm intra} = U^{bb}_{\rm bond} + U^{bb}_{\rm angle} + U^{bb}_{nb}.
\end{equation}

The intramolecular bond energy
\begin{equation}
\label{ubond}
U^{bb}_{\rm bond} = U_{\rm Ed}^{bb} +  \sum_{i=1}^{n} \sum_{\alpha=1}^{n_b-1}
v_{\rm rep}^{bb}(l_{i\alpha}),
\end{equation}
where the first term is given by Eq.\eqref{edwardsH} above, and the pair potential, $v_{\rm rep}^{bb}(r)$, provides a short-range repulsion between nearest neighbor blobs. It will be defined below.

The intramolecular angle energy is defined as
\begin{equation}
\label{uangle}
    U^{bb}_{\rm angle} = -k_{\rm B}T\sum_{i=1}^n \sum_{\xi=1}^{n_b-2} \ln[P(\theta_{i\xi})/\sin(\theta_{i\xi})]
\end{equation}
where $\theta_{i\xi}$ is the bond angle between the sites $\xi, \xi+1$ and  $\xi+2$. the angular probability distribution is given for a chain following a random walk as
\begin{equation}
\begin{split}
    P(\theta) = & \frac{(1-a^2)^{3/2} \sin{\theta}}{\pi (1-a^2\cos^2{\theta})^2} \\
    & \Big[\frac{1+2a^2 \cos^2{\theta} \arccos{(-a \cos{\theta})}}{\sqrt{1-a^2 \cos^2{\theta}}} + 3 a \cos{\theta}\Big]
\end{split}
\end{equation}
with $a \rightarrow -0.25$ for long chains.\cite{Laso1991}

The non-bonded, long-range intramolecular energy, $U^{bb}_{nb}$, is represented as a sum of pair-wise interactions given by the effective IECG intermolecular potential, $U^{bb}_{\rm eff}(r)$, formally derived in previous publications,\cite{Clark2012,Clark2013,Clark2015} and discussed below.

The intermolecular IECG internal energy, $U^{bb}_{\rm inter}(\mathbf{R}_n)$, is a sum of pairwise interactions between CG sites on different chains:
\begin{equation}
U^{bb}_{\rm inter}(\mathbf{R}_n)=\sum_{i=1}^{n-1}\sum_{j>i}^n \sum_{\alpha=1}^{n_b} \sum_{\gamma=1}^{n_b} U^{bb}_{\rm eff}(r_{i\alpha j\gamma}),
\end{equation}
where $r_{i\alpha j\gamma} \equiv \vert \mathbf{r}_{i\alpha}-\mathbf{r}_{j\gamma}\vert$.

The effective potential, $U^{bb}_{\rm eff}(r)$, is determined by inverting a physically reasonable closure to the PRISM equation. For descriptions of molecular liquids at the site level with relatively short-range interactions, a closure to the PRISM equation\cite{Schweizer1997} that accounts for correlations at the molecular level appears to be necessary.\cite{Schweizer1997,Donley1994} However,  the IECG potential, $U^{bb}_{\rm eff}(r)$, is long-ranged and an atomic closure that models well long-range interactions is sufficient. In this case, the hypernetted-chain (HNC) closure serves this purpose well. Inverting that closure yields
\begin{equation}
\label{ueff}
U^{bb}_{\rm eff}(r) =-k_{\rm B}T\left [ \ln{(g^{bb}(r))} - h^{bb}(r) + c^{bb}(r) \right ],
\end{equation} 
and so the CG potential depends on the intermolecular total correlation function $h^{bb}(r)$ between blobs, where $g^{bb}(r)=h^{bb}(r)+1$, and the blob direct correlation function, $c^{bb}(r)$.

The repulsive pair potential appearing in Eq.\eqref{ubond} is chosen to be Eq.\eqref{ueff}, but with the potential of mean force subtracted:
\begin{eqnarray}
v_{\rm rep}^{bb}(r) &=& U^{bb}_{\rm eff}(r)  + k_{\rm B}T \ln{(g^{bb}(r))} \nonumber \\
&=& k_{\rm B}T \left [ h^{bb}(r) - c^{bb}(r) \right ]. 
\end{eqnarray}
The second expression is the negative of the HNC medium-induced potential and so is repulsive at short distances as required.

The above forms for $U_{\rm bond}^{bb}$ and $U_{nb}^{bb}$ make the CG intramolecular energy dependent on density and temperature through $h^{bb}(r)$ and $c^{bb}(r)$. Since this energy is used to compute the partition functions $Z_n^{bb}$ for any $n$, in the IECG theory one selects the density of the true liquid. In that way, the partition function of a single isolated chain, $Z_1^{bb}$, depends on the density of the liquid into which the chain will be inserted.

An appropriate PRISM equation relates $c^{bb}(r)$ to $h^{bb}(r)$.  In Fourier space it is 
\begin{equation}
\label{cbb}
\hat{c}^{bb}(k)=\hat{h}^{bb}(k)/  [n_b \hat{\Omega}^{bb}(k)[ \hat{\Omega}^{bb}(k) + \rho_b \hat{h}^{bb}(k)  ] \ ,
\end{equation} 
where a caret denotes the Fourier transform of its respective quantity, with $k$ being the wavevector, and $\hat{\Omega}^{bb}(k)$ is the blob-blob intramolecular structure factor, i.e., Fourier transform of the intramolecular pair distribution function (pdf). The effective potential is then uniquely specified knowing $h^{bb}(r)$ .

This intermolecular total correlation function between CG sites, $h^{bb}(r)$, can be computed from the monomer-monomer correlation function, $h^{mm}(r)$, using an extension to liquids of multiblob chains of an equation derived by Krackoviak, Hansen and Louis (KHL) for polymers in solutions represented as soft spheres.\cite{krakoviack2002relating}  To do this, KHL divide the sites on the chain into real ones and  fictitious ones, the latter in our case being CG units (blobs). They then adopt a Chandler-Andersen interpretation of the direct correlation function, which means it behaves as an effective potential. In that way, the direct correlation between a blob and any other site (monomer or blob) is approximated as zero. The KHL equation in Fourier space becomes in our case\cite{Clark2012,Clark2013,Clark2015}  
\begin{equation}
\label{hbb}
\hat{h}^{bb}(k)=\hat{\Omega}^{bm}(k)^2 \hat{h}^{mm}(k)/ \hat{\Omega}^{mm}(k)^2 \ .
\end{equation}

Since Eq.\eqref{hbb} relies on the knowledge of $h^{mm}(r)$ at long wavelengths, it might seem that determining $h^{bb}(r)$ would necessitate extending simulations over very long times, a situation best avoided. However, as discussed below, the IECG theory employs an alternative approach that only requires short-time information, such as for the pressure.

The intramolecular structure factors in Eqs.\eqref{cbb} and \eqref{hbb} for a flexible chain are all appropriately normalized, and denoted as $\hat{\Omega}(0) = 1$.\cite{Clark2012,Clark2013,Clark2015} The monomer-monomer structure factor  can be reasonably approximated using the Debye function\cite{Clark2013}
\begin{equation}
\hat{\Omega}^{mm}(k)=\frac{2}{n_b^2 q_b^2} (n_b q_b -1 + e^{-n_b q_b}) \ ,
\label{Debye}
\end{equation}
where $q_b=q/n_b$ and $q=k^2 \langle R^2\rangle /6$. And the blob-monomer one is similarly found to be 
\begin{eqnarray}
\hat{\Omega}^{bm}(k) & = &\frac{1}{n_b} \Bigg[\frac{\sqrt{\pi}}{q_b^{1/2}}      \Bigg] \text{erf}\left(\frac{q_b^{1/2}}{2}\right)e^{-q_b/12} \nonumber \\
- & 2& \Bigg(\frac{e^{-n_bq_b}-n_be^{-q_b}+n_b-1}{n_bq_b(e^{-q_b}-1)e^{-q_b/3}} \Bigg) \ , 
\label{WCMK}
\end{eqnarray}
while the blob-blob one is \begin{equation}
\label{Wbb}
\hat{\Omega}^{bb}(k)=\frac{1}{n_b} + 2 \Bigg(\frac{e^{-n_bq_b}-n_be^{-q_b}+n_b-1}{n_b^2(e^{-q_b}-1)^2} \Bigg) e^{-2q_b/3} \ .   
\end{equation}

In liquids of long polymer chains, we observe a very high level of accuracy in the agreement between the numerical and analytical solutions for these intramolecular structure factors.\cite{Dinpajooh2018}

Lastly, the total correlation function for monomer-monomer interactions in Eq.\eqref{hbb} is determined by the monomer-monomer PRISM equation.\cite{schweizer_prism_1994} In Fourier space, this equation can be expressed as 
\begin{equation}
\label{prism_mm}
\hat{h}^{mm}(k)=\hat{\omega}^{mm}(k)^2 \hat{c}^{mm}(k)/[1-\rho \hat{\omega}^{mm}(k) \hat{c}^{mm}(k)]\ .
\end{equation} 
Here $\hat{\omega}^{mm}(k)=N\hat{\Omega}^{mm}(k)$, with the latter being defined  in Eq.\eqref{Debye} above. Determining the Fourier transform of the direct correlation function, $\hat{c}^{mm}(k)$, is the final step in completing the theory. 

In previous studies of the IECG theory for long polymer chains, it has been observed that $\hat{c}^{mm}(k)$ can be accurately approximated by just its $k=0$ value, $\hat{c}^{mm}(k=0)=c_0$. This can be attributed to the fact that the CG effective potential exhibits a significantly longer range than the actual monomer-monomer interaction potential. Consequently, also the direct correlation functions, who are functions of their potentials, also follow a similar pattern. Thus, the full IECG potential, $U^{bb}_{\rm inter}$, which depends on the monomer direct correlation function through the equations above, is defined after the fitting of a single adjustable parameter, $c_0$ (for more details, please refer to the Supplementary Material (SM)). 

Within the IECG theory, the value of $c_0$ is, in turn, derived from the knowledge of the pressure. For high-density systems comprising long polymer chains, the HNC closure can be effectively approximated by the Mean Spherical Approximation (MSA) closure. In this context, Eq.\eqref{ueff} is replaced by $U^{bb}_{\rm eff}(r) = -k_{\rm B}T c^{bb}(r)$. From this, the pressure in the IECG theory can be expressed as\cite{Clark2013,Dinpajooh2018}
\begin{equation}
\label{pressure}
    \frac{P}{\rho_{ch}k_{\rm B} T} = 1-\frac{\rho N c_0}{2} \ .
\end{equation}

This CG pressure is subsequently equated with the pressure of the actual system, thereby determining the value of $c_0$. Through this parameter, the characteristics of the CG system, including the energies $U^{bb}_{\rm intra}$ and $U^{bb}_{\rm inter}$, are established once the pressure of the actual system, which is a quantity that rapidly reaches equilibrium in a simulation, is determined.

\subsection{Chemical potential in the IECG theory}
Let's begin by considering the ideal contribution to the chemical potential, as given in Eq.\eqref{mu_id}. By substituting Eq.\eqref{Zbb1} into \eqref{mu_id}, we can observe that $\mu_{\rm id}$ appears to depend strongly on the degree of coarse-graining through $n_b$. This is the case even when we take into account that the average bond length $l$ tends to decrease as $n_b$ increases. This behavior is expected to persist even for more realistic chain models.

Note however that in determining phase coexistence, the chemical potentials of the two phases must be equal. Therefore, it is the difference between the chemical potentials of the two phases that matters. To address this, we introduce a reference single-chain intramolecular energy, designated as $U_{0}^{bb}$, and define the difference between this reference and $U_{\rm intra}^{bb}$ for $n=1$ as $\Delta U_{\rm intra}^{bb}$. With these definitions in place, we can reframe Eq.\eqref{mu_id} using a Widom-like approach as
\begin{equation}
\label{mu_idWidom}
\mu_{\rm id} = -k_{\rm B}T\ln \left [\frac{ \langle e^{-\beta\Delta U_{\rm intra}^{bb}} \rangle_0 Z_0}{\Lambda^{3n_b} n}\right ] \ .
\end{equation}
Here, the average ($\langle \rangle_0$) is taken with respect to $U_0^{bb}$, while $Z_0$ denotes the partition function of a single chain in the reference system. The calculation of this single-chain average can be achieved through simulations. Since $Z_0$ remains (approximately) the same for both phases, it can be set to $V\zeta$, with the specific value of $\zeta$ being disregarded. Consequently, all the relevant terms in $\mu_{\rm id}$ can be determined in a straightforward manner.

Now, let's turn our attention to the excess chemical potential, as defined in Eq.\eqref{mu_exc}. To simplify this equation, we follow a traditional approach proposed by  Hansen\cite{Hansen1977} and Verlet \cite{Verlet} for simple liquids. The objective is to reduce the equation to a form that depends solely on the quantities defined in Sec. \ref{IECGformalism} above.

Following the conventional procedure for simple liquids, we regard the intermolecular energy, $U^{bb}_{\rm inter}$, as a smoothly varying function of a switching parameter, $\lambda$, which characterizes the coupling of an inserted particle with its surrounding matrix.
When $\lambda=1$, the system consists of $n$ molecules, including the inserted one, fully interacting, while $\lambda=0$ represents a scenario where the inserted molecule does not interact with the remaining $n-1$ molecules.\cite{Hill1987} 

With this in mind, Eq.\eqref{mu_exc} can be expressed as  
\begin{equation}
\label{uu}
\mu^{bb}_{\rm exc}=-k_{\rm B}T \int_0^1d\lambda \ \frac{\partial \ln{ Z^{bb}_n(\lambda)}}{\partial \lambda} 
\end{equation}

Let's assume that there is no change in the intramolecular structure of the inserted CG chain as the interaction parameter $\lambda$ is gradually increased. Under this assumption, the excess potential in the HNC approximation for $U^{bb}_{\rm eff}(r)$, as given in Eq.\eqref{ueff}, adopts the form proposed by Hansen\cite{Hansen1977} and Verlet\cite{Verlet}:
\begin{equation}
\label{mu_theory}
\mu^{bb}_{\rm exc}  \approx   k_{\rm B}T n_b \rho_b \int d \textbf{r} \    \left [\frac{1}{2} \  h^{bb}(r)[h^{bb}(r)-c^{bb}(r)]-c^{bb}(r)\right ]  .
\end{equation}

Furthermore, when applying the MSA closure (that can be derived from the HNC closure by assuming that the pair distribution function is equal to $1$), Eq.\eqref{mu_theory} simplifies to
\begin{equation}
\label{analytical}
\begin{split}
    \mu_{\rm exc}^{bb} \approx - k_{\rm B} T n_b \rho_b \int d \textbf{r} \ c^{bb}(r) = -N k_{\rm B} T \rho c_0  \ .
\end{split}    
\end{equation}

It is important to note that within these approximations, both the pressure and the excess chemical potential remain invariant in relation to the degree of coarse-graining. 
Interestingly both Eq. (22) and Eq.(26) do not depend on the number of blobs, $n_b$.
 This property proves to be valuable when employing the IECG model to construct simulations that encompass multiple scales for a polymer liquid, as it allows the flexibility to fine-tune the simulation resolution as needed.

In the context of a CG polymer modeled as a single soft sphere, where the intramolecular contribution to the excess chemical potential is null, Eq.\eqref{mu_theory} and \eqref{analytical} are expected to be very accurate. In a multiblob representation, the excess chemical potential will also include an intramolecular contribution, which  incorporates the effect of chain structure change from the ideal to melt state. For a long polymer chain partitioned in a small number of soft blobs, the influence of excluded volume effects is expected to be minimal, so that intramolecular contribution to $\mu^{bb}_{\rm exc}$ can be safely neglected. However, as we progressively divide the polymer into an increasing number of CG sites, each representing shorter and shorter segments, this approximation is expected to become progressively less reliable. 

But are these expectations correct? That is, if one computed $\mu_{\rm exc}^{bb}$ from a realistic simulation of the CG system, would it be invariant at least up to a modest number of CG units, say $n_b \leq 6$? Also, would its value be well approximated by Eq.\eqref{mu_theory} or \eqref{analytical}? In the following sections, we present evidence to address these questions.

\section{Methods}
\subsection{IECG simulations}
\label{MCCCS-CG}
In this section we provide a comprehensive description of the coarse-grained simulation methodology employed to validate the theoretical framework.
Siepmann and coworkers have developed an open-source software package named     that performs Gibbs ensemble Monte Carlo for Complex Chemical Systems (MCCCS) for performing Gibbs Ensemble Monte Carlo (GEMC) simulations of polymers and other molecular systems. This software incorporate the gradual chain insertion technique using the configuration-biased Monte Carlo (CBMC) algorithm.  The code has been further extended and equipped with a user-friendly interface by Martin in a version called MCCCS Towhee.\cite{Martin2013} Initially, we extended the MCCCS code to compute the excess chemical potential for polymer liquids in the united atom representations, employing the CBMC methodology. Then we included the IECG model to calculate the excess chemical potential of coarse-gained polymers. We modified an old version of the MCCCS code from 2010, V2.2. 

The modified code, herein referred to as ``MCCCS-CG", is employed to perform Monte Carlo simulations of the IECG liquid. It is used to calculate the energy variation and the corresponding change in chemical potential resulting from the insertion of a single CG chain. 

The MCCCS-CG code differs from the original MCCCS V2.2 in three minor ways: 1) all pairwise potentials such as $U^{bb}_{\rm eff}(r)$, Eq.\eqref{ueff}, are computed on a grid of points in $r$ and read from a file. 2) to perform chain insertion the CBMC algorithm is not used, but we use a simpler algorithm based upon the expressions for $U^{bb}_{\rm inter}$ and $U^{bb}_{\rm intra}$ discussed in Sec. \ref{IECGformalism} above. In this algorithm different conformations of polymer chains are added all at once since the potentials are soft and the number of blobs is small, and the probability of insertion of the whole IECG chain in one step is found to be high, even in high-density melts. Lastly, 3) a simple Widom expression is used to compute the excess chemical potential.

This Widom expression is merely Eq.\eqref{mu_exc}, which can be rewritten as
\begin{equation}
\label{mu_exc_sim}
 \mu_{\rm exc}^{bb} = - k_{\rm B} T \ln \left \langle \exp{\left(-\beta\Delta U^{bb}_{\rm inter}\right)} \right \rangle\ .
 \end{equation}
In this context, $\Delta U^{bb}_{\rm inter}$ represents the interaction energy of the inserted particle and the existing polymer matrix. The brackets ($\langle\ \rangle$) denote an average over the configurations of the inserted chain and the preexisting ones. To calculate this value, a chain is introduced into the polymer matrix, and the change in internal energy resulting from the insertion is determined. Upon repeated insertions and equilibration of the polymer matrix itself, the average in Eq.\eqref{mu_exc_sim} is then obtained. 

In the initial stage, we selected a cubic box with periodic boundary conditions and prepared an equilibrated polymer matrix through Monte Carlo simulations. To attain equilibrium of the IECG multiblob liquid, the MCCCS-CG code implements three distinct types of moves in each cycle: rotation of a single-chain around its center-of-mass, linear translation of the center-of-mass, and translation of individual CG sites. When representing the polymers as soft-spheres, the only move performed is the translation of the center-of-mass. Each move is attempted on a randomly selected chain and repeated a number of times equal to the total number of polymers within the sample. Translational moves of each IECG site are attempted $50 \ \%$ of the times, molecule translations are tried $40 \ \%$ of the times, and a molecular rotations are attempted $10 \ \%$ of the times.

Following the completion of an entire equilibration cycle, the MCCCS-CG code initiates thousands of attempts to insert complete polymer chains. The code selects a conformation for insertion from the existing matrix's population in a random manner. As a result of this process, the distribution of conformation for the inserted chain statistically complies with the correct Boltzmann distributions for bonds and angles. The chain's conformational statistic adhere to an unperturbed distribution prior to insertion.

Remarkably, we observe that the convergence of $\mu_{\rm exc}^{bb}$ is quite rapid when utilizing the MCCCS-CG code. Convergence is typically achieved after approximately $10^5$ insertions for soft spheres and up to $10^7$ insertions for chains represented by ten blobs. The number of steps required for convergence tends to increase as the number of monomers within each blob decreases, primarily because the IECG potential exhibits a sharper profile.

To expedite the computation, we introduced a cut-off distance for the intermolecular potential and incorporated the necessary tail corrections when recording the excess chemical potential, following the approach employed in the original MCCCS V2.2 code.

\section{Results}
\subsection{An example of MCCCS-CG calculations: the excess chemical potential as a function of density and temperature for a polymer liquid with a degree of polymerization, N=44}
As an initial illustration of the method we use in MCCCS-CG simulations, we present calculations concerning the excess chemical potential for high-density melts composed of relatively short polymer chains, with $N=44$ monomers. This is a scenario where the insertion procedure rapidly achieves convergence. 

\begin{figure}[htb]
\includegraphics[width=.9\columnwidth]{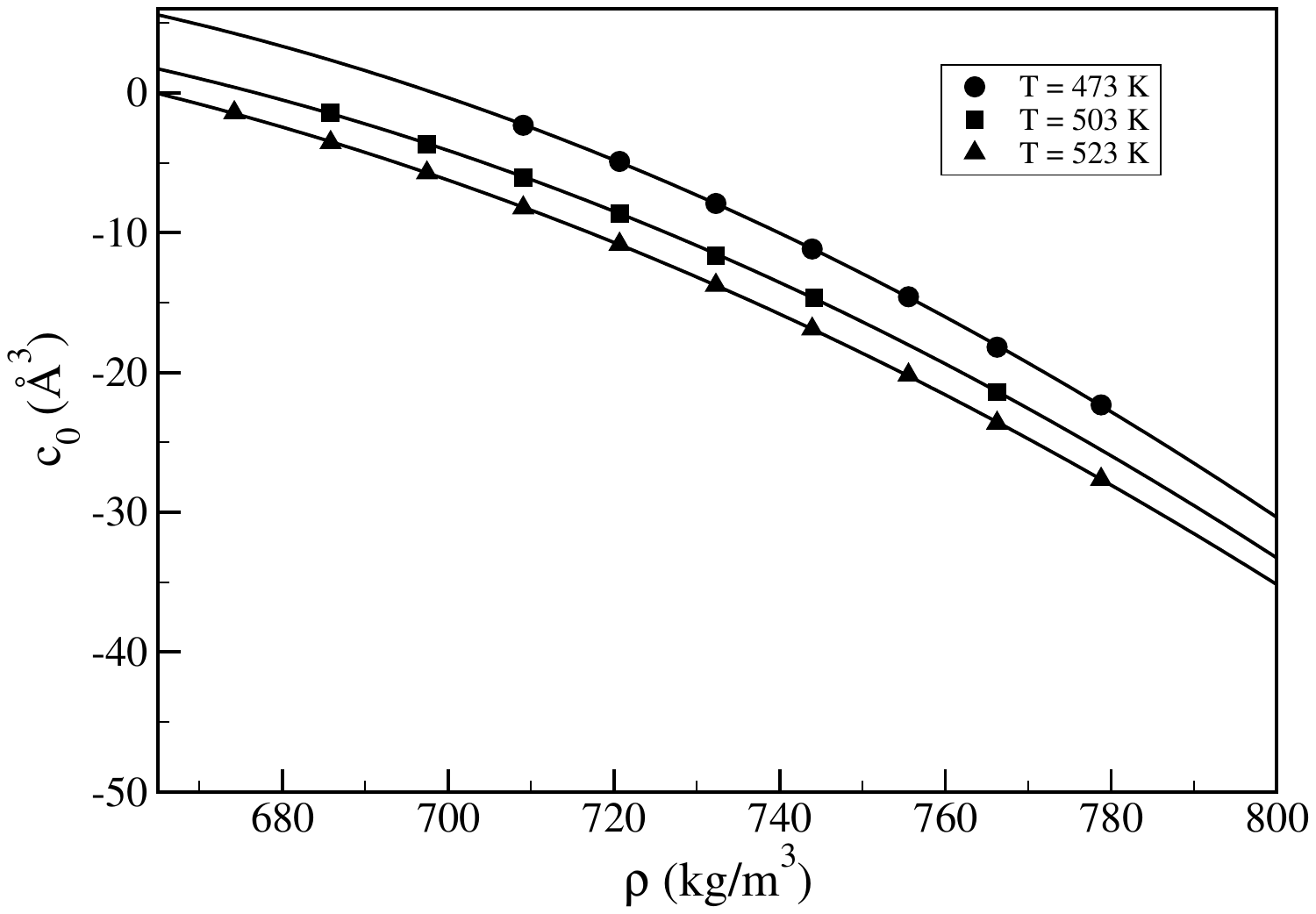}
\caption{The values of $c_0$  computed from atomistic simulations of polymer melts of degree $N$ = 44 at increasing density and at different temperature. Data are fitted by equations that allow one to extrapolate the parameters to polymer melts of other densities and temperatures. A similar functional expression effectively interpolates the $c_0$ data across various degrees of polymerization (see the SM for details)}
\label{C0figure1}
\end{figure}

The force field input to the MCCCS-CG simulations that calculate the excess chemical potential is the CG IECG forcefield described in Sec.\ref{IECGformalism} above, which depends on the monomer-monomer direct correlation function, $c_0$, through the HNC closure. As mentioned there, $c_0$ is determined by equating the CG pressure with that of the true system, which here we take to be an atomistic molecular dynamics (MD) simulation of the same system. The LAMMPS atomistic simulation software\cite {Plimpton1995} was used.

\begin{figure}[tbh]
\includegraphics[width=0.72\columnwidth]{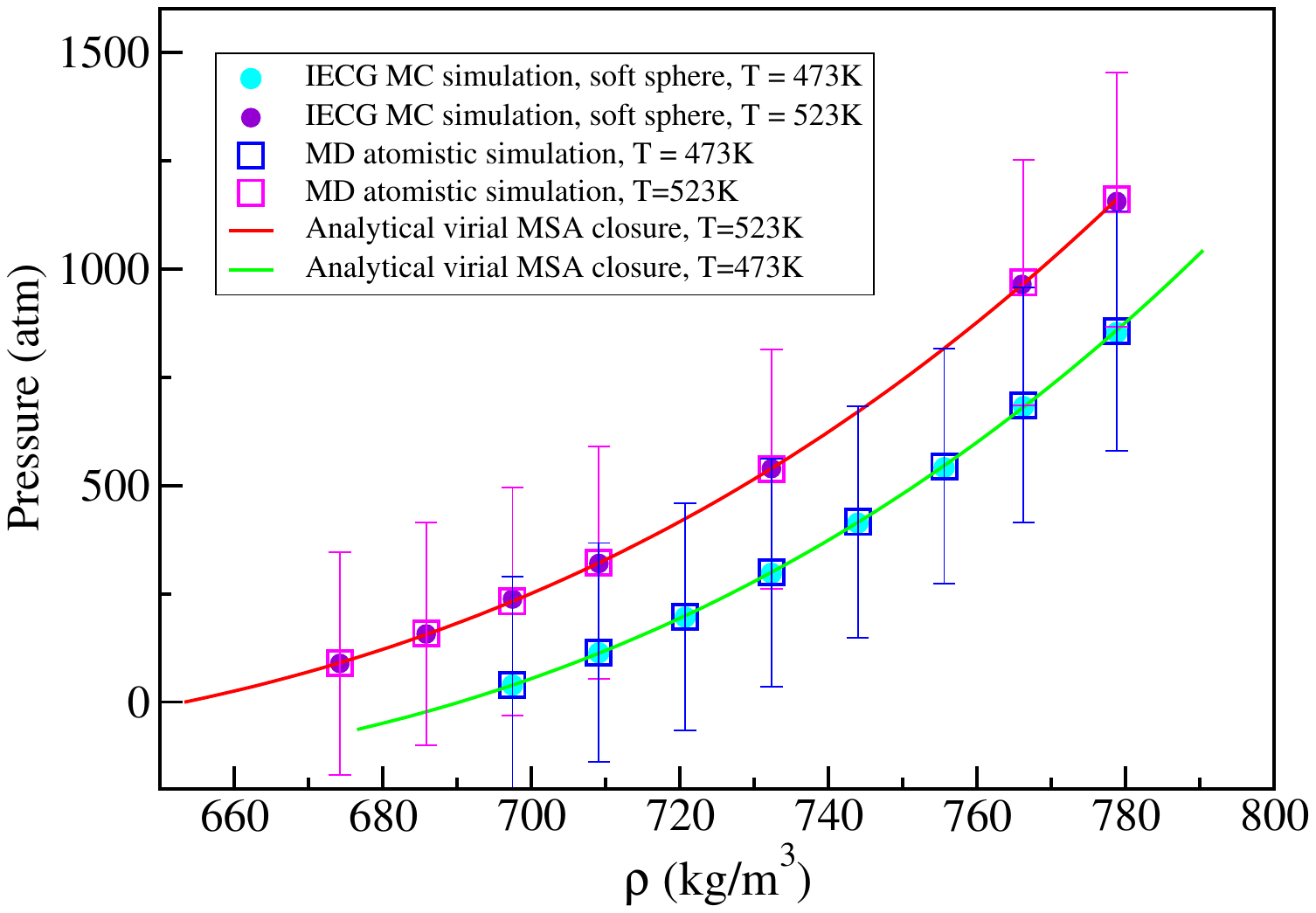}
\caption{Comparison of the virial pressure measured in the MCCCS-CG simulation for PE melts with $N=44$ and chains represented by one soft sphere with the virial pressure measured in atomistic MD simulations performed with the LAMMPS code. The soft sphere IECG simulations are performed at temperatures $T=473 \ K$ (blue circles)  and $T=523 \ K$ (violet circles). The atomisitc LAMMPS simulations are performed at the same temperatures, $T=473 \ K$ (blue squares) and $T=523 \ K$ (pink squares). For the atomistic simulations, we also report the variance of the measured pressure. In contrast, the variance of the pressure in the CG simulation is smaller than the size of the symbols. The pressure agreement between the CG MC and the atomistic MD simulations is quantitative at both temperatures. The analytical solution of the virial pressure with our model and the mean-field MSA closure at $T=473 \ K$ (green line) and $T=523 \ K$ (orange line) also shows quantitative agreement.}
\label{MCOPressure}
\end{figure}

Figure \ref{C0figure1} shows the optimized $c_0$ parameter for a series of simulations at three distinct temperatures and increasing densities.  The optimized parameter traces a seamless curve that can be conveniently fitted by an empirical equation as a function of density and temperature. A similar functional expression effectively interpolates the $c_0$ data across various degrees of polymerization (see the SM for more details).
By extrapolating these empirical equations, we can calculate the IECG potential for novel state points, obviating the necessity for new atomistic simulations.  Consequently, the IECG potential proves transferable to diverse polymer liquids operating under different thermodynamic conditions.

\begin{figure}[tbh]
\includegraphics[width=0.8\columnwidth]{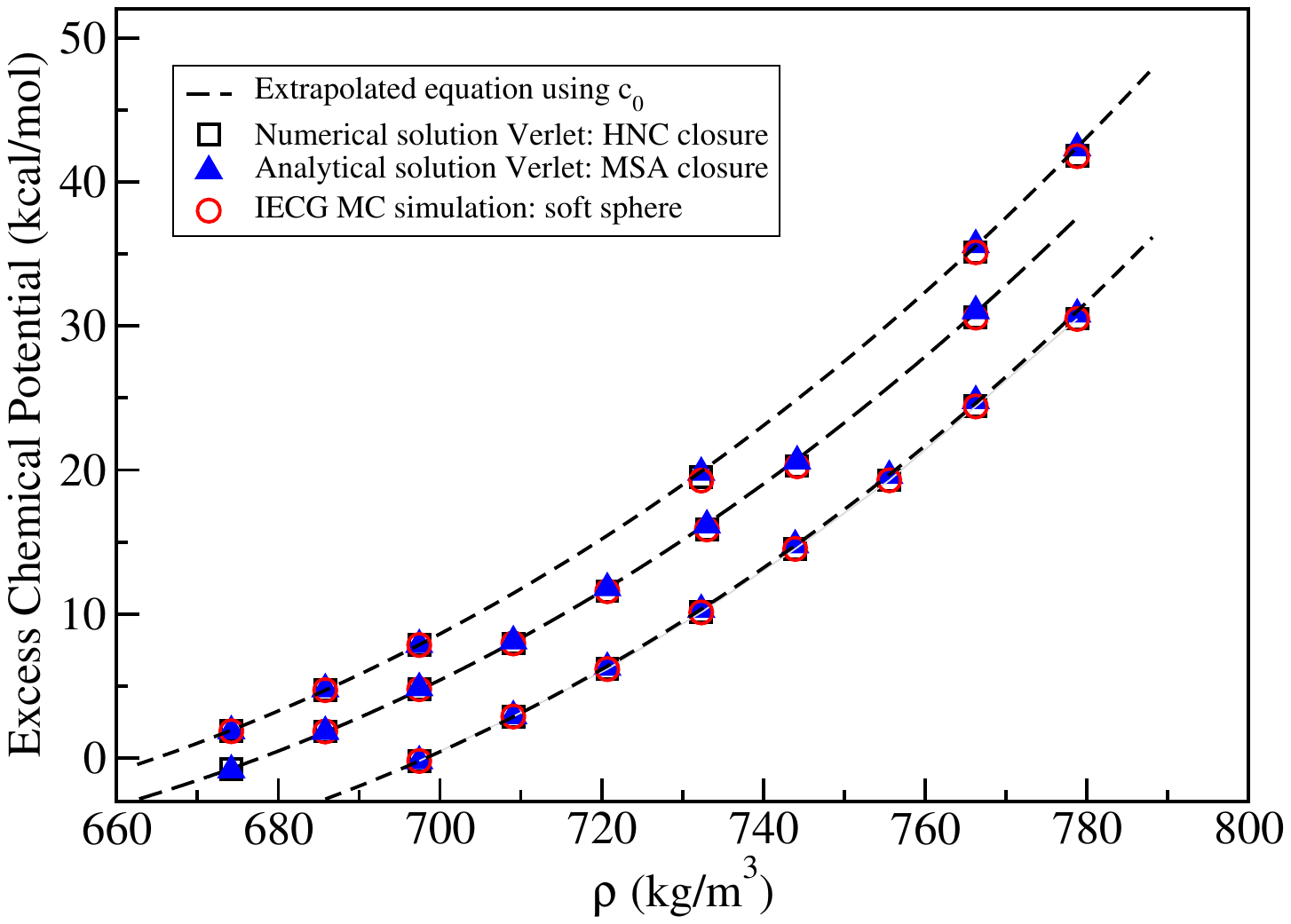}
\caption{Excess chemical potential, $\mu_{\rm exc}^{bb}$, as a function of density, $\rho$, for a PE melt with $N=44$ for three different temperatures (from the top to the bottom: $T=473 \ K$, $503 \ K$, and $523 \ K$). Theory predictions using the HNC closure, Eq.\eqref{mu_theory} are shown as black squares, predictions using the MSA closure, Eq.\eqref{analytical} are shown as solid blue triangles, and MCCCS-CG simulation data are shown as red circles. The polymers are modeled as soft spheres, so $n_b=1$. The black dashed line corresponds to predictions using the MSA closure with extrapolated data of $c_0$ shown in  Fig. \ref{C0figure1}.}
\label{MCON44}
\end{figure}

Subsequently, equipped with the solved IECG potential as input, we executed MCCCS-CG simulations on the CG polymer liquid. To assess the accuracy of the MCCCS-CG simulation in establishing a suitably equilibrated liquid matrix for chain insertion, our initial step involved verifying that the structural and thermodynamic properties of the IECG melt, as predicted by MCCCS-CG code, align with the properties observed in an equivalent atomistic MD simulation. The details of the MD simulation are reported in the  SM.

Figure \ref{MCOPressure} shows the pressure as a function of density computed from the MCCCS-CG and atomistic LAMMPS simulations at various thermodynamic conditions. As can be seen, results of the two types of simulations are in excellent agreement, even though the variance in the atomistic simulation data is much larger than for the CG for the same run time. This agreement indirectly validates the analytical approximation to the pressure, Eq.\eqref{pressure}, including the use of the MSA closure.

It is important to note that Figure \ref{MCOPressure} selectively presents pressure data for the highest and the lowest temperatures simulated. This approach is taken to prevent undue clutter, as including intermediate temperature data could overly complicate the visualization. However, it is worth highlighting that a comparable level of agreement is observed for the intermediate temperature as well.
The MCCCS-CG and atomistic LAMMPS simulations were also found to be in agreement for the pair distribution function, $g^{bb}(r)$, as reported in the SM.
As mentioned above, in the atomistic simulation the pressure equilibrates quickly. Consequently, the determination of $c_0$ only necessitates a relatively short MD trajectory of a few nanoseconds (less than 3 $ns$ even for the longest chains considered in this study, $N=720$). 

Figure \ref{MCON44} shows the excess chemical potential as function of density for three temperatures, $T=473$ K, $503$ K, and $523$ K. Data from the MCCCS-CG simulation, the numerical solution of the HNC expression, Eq.\eqref{mu_theory}, and the analytical MSA solution, Eq.\eqref{analytical}, are shown. As anticipated, the chemical potential increases with increasing density since inserting the chain becomes more challenging with greater crowding. Increasing the temperature though facilitates chain insertion, thereby reducing $\mu_{\rm exc}^{bb}$, as one would expect. As can also be seen, the numerical and analytical values are in excellent agreement with each other, and with the simulation data.

Additionally, we subjected the MSA equation governing the excess chemical potential to testing using the empirical equation that characterizes the density-dependent behavior of the direct correlation function, $c_0$, (see Figure \ref{C0figure1}). Notably, the resulting dashed line depicted in Figure \ref{MCON44} precisely intersects the numerical values of the recently computed excess chemical potential.  This agreement highlights that the MSA approximation stands capable of offering a dependable initial estimate for the excess chemical potential within high-density polymer melts.

Figure \ref{MCOnb300} confirms this invariance by showing the behavior of $\mu_{\rm exc}^{bb}$ for a longer chain, $N=300$. Across different temperatures and densities, this invariance remains up to a partition of the chain in $n_b=10$ CG sites, where each site represents a $N_b=30$ monomer segment.
Again, simulation data and theory predictions are in excellent agreement.  Throughout both figures, any notable deviations from the constant trend emerge under the conditions of lowest temperature and highest density, where the insertion of the chain becomes less favorable.

As anticipated, the invariance in the IECG model is robust, albeit it diminishes as the coarse grained chain is divided into shorter and shorter segments. Specifically, predictions with CG sites that include $N_b=30$ monomers differ slightly from those of larger $N_b$ (smaller $n_b$). Under such conditions, the distribution of monomers within the chain starts deviating from Gaussian statistics. This invariance stands as a valuable trait of the IECG model, particularly in the context of multiscale modeling procedures, ensuring that the granularity of the CG model can be flexibly adjusted without introducing undesirable spurious forces.

\subsection{The excess chemical potential is independent of the  granularity of the model}
The analytical solution of the excess chemical potential under conditions of high density and low granularity, Eq.\eqref{analytical}, indicates that it remains unaffected by the degree of coarse-graining, $n_b$, in the IECG model. However, as the conditions required for the application of the MSA closure cease to be met, the invariance of $\mu_{\rm exc}^{bb}$ with $n_b$ becomes questionable. In this section, we explore the steadfastness of the analytical approximation of $\mu_{\rm exc}^{bb}$ by varying the CG granularity in the chain model.

The accuracy of the IECG theory to estimate the excess chemical potential at various levels of coarse-graining  is first validated by comparing the predictions of Eq.(\ref{mu_theory}) with MCCCS-CG simulation data that use Eq.(\ref{mu_exc_sim}). We examined a polymer melt with degrees of polymerization of $300$ at a density of $\rho= 766$ kg/m$^{3}$. 

The results from the MCCCS-CG code are listed for various levels of coarse-graining in Table \ref{tablea}. This Table displays the calculated values of the excess chemical potential along with their corresponding tail corrections. It is evident that these tail corrections are indispensable for ensuring the accuracy of the chemical potential calculations. Once these tail corrections are incorporated, the excess chemical potential values from IECG simulations become nearly independent of the model's granularity. 

The right side of Table \ref{tablea}  also presents the chemical potential values derived from the IECG theory, demonstrating excellent agreement with the excess chemical potential obtained from the MCCCS-CG simulations. Therefore, both IECG simulations and IECG theory indicate that the excess chemical potential remains unaffected by the specific CG model adopted, as long as the IECG polymer coarse-graining approach is utilized.

It is worth noting that, to the best of our knowledge, direct comparisons between the results of the chemical potential for melts of long polymer chains from the IECG theory or simulations and atomistic simulations are currently computationally unfeasible. However, future work will explore the exciting possibility of combining the IECG method with advanced MC algorithms at the atomistic level, which could pave the way for such calculations, as discussed later. 

\begin{table}[h!]
\begin{center}
\caption{Comparison of the values of the excess chemical potential obtained from the IECG MC simulations (Eq.(\ref{mu_exc_sim}): $\mu_{\rm exc,sim}$)  and the IECG theory (Eq.(\ref{mu_theory}): $\mu_{\rm exc,theory}$) for melts of polyethylene chains with degree of polymerization, $N=300$. The table also reports the tail corrections for the calculations of the excess chemical potential from IECG simulations. The density of polymer melt is $\rho= 766$ kg/m$^3$, and their  temperature is $T= 503$ K. The systems have variable cut-off distance, $r_{\rm cut}$, variable box length, $l_x$, and variable number of molecules, $n$. Each polymer is modeled with a variable number of blobs, $n_b$. All the simulations were performed with the MCCCS-CG code. To the best of our knowledge, it is currently computationally intractable to obtain the excess chemical potential at the atomistic resolutions for these systems.}
\label{tablea}
\begin{tabular}{ c | c | c | c | c| c | c }
\hline
\hline
$n_b$ &  $n$ & $l_x$  & $r_{\rm cut}$ & $\mu_{\rm tail}$ & $\mu_{\rm exc,sim}$&  $\mu_{\rm exc,theory}$\\
&  & \AA  &   \AA  &  kcal/mol & kcal/mol &  kcal/mol \\
\hline
\hline
$1$ & $20258$ & $569.17$ & $210.8$ &    $-8.1$& $89.0\pm0.5$ & $89.0$ \\
$2$ & $6211$  & $383.8$  & $127.9$ &   $-6.7$ & $88.4\pm0.7$ & $88.6$ \\
$3$ & $2610$  & $287.5$ & $95.8$ &   $-6.7$ & $87.9\pm0.9$ & $88.1$ \\
$4$ & $1416$ & $234.5$  & $78.2$ &   $-5.1$ & $87.2\pm1.1$ & $87.8$ \\
$6$ & $799$  & $193.7$ & $58.7$ &   $-4.1$ & $86.4\pm1.5$  & $87.2$ \\
$10$ & $300$ &  $139.8$  & $40.9$ &   $-2.8$ & $86.2\pm2.0$ & $86.8$ \\
\hline
\end{tabular}
\end{center}
\end{table}

\begin{figure}[tbh]
\includegraphics[width=0.75\columnwidth]{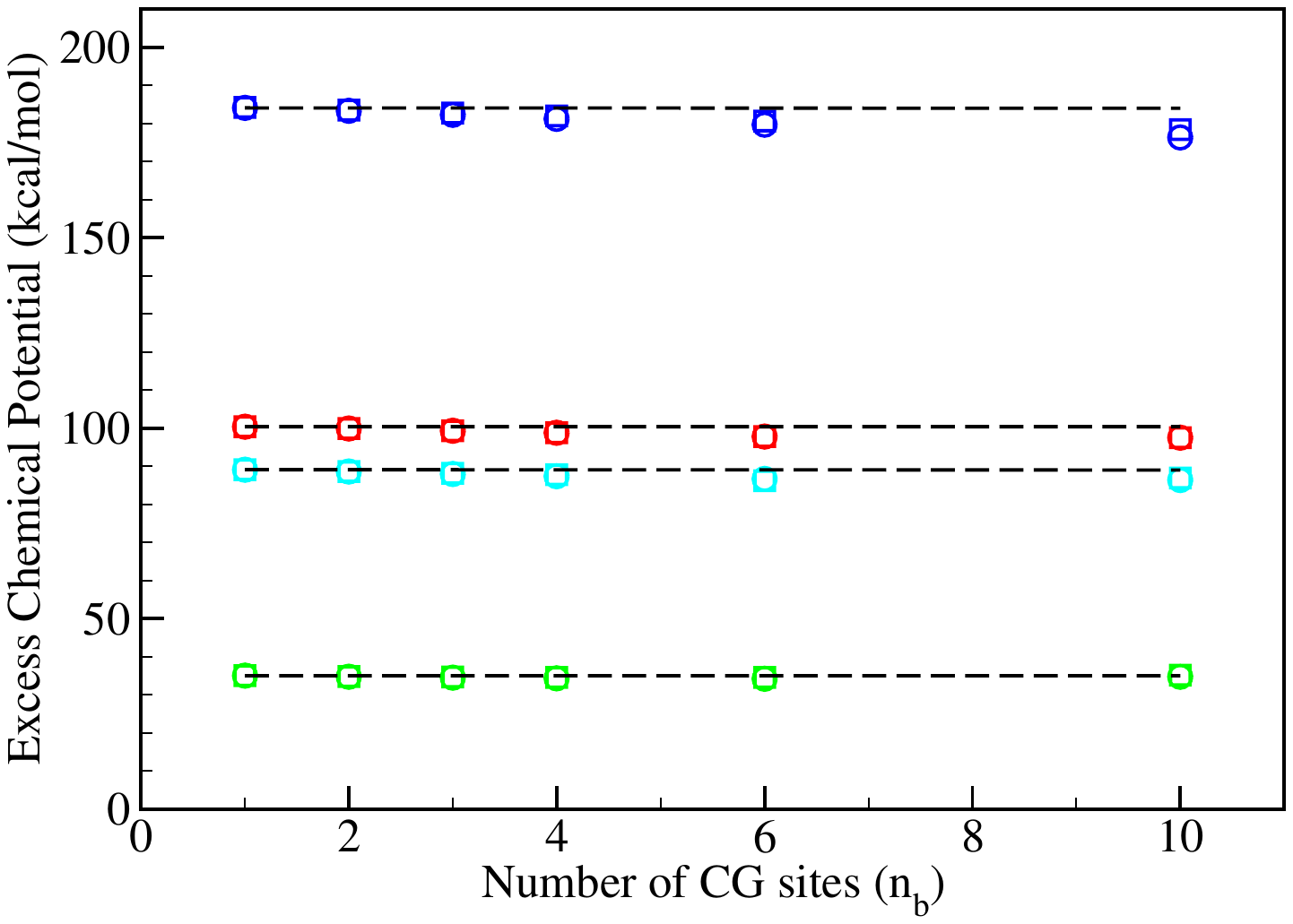}
\caption{Excess chemical potential for a melt of PE chains of degree of polymerization, $N=300$, at variable temperature and density. From the top to the bottom: blue symbols $T=503$ K and $\rho=797$ kg/m$^{3}$; red symbols $T=513$ K and $\rho=766$ kg/m$^{3}$; cyan symbols $T=503$ K and $\rho=766$ kg/m$^{3}$; green symbols $T=503$ K and $\rho=744$ kg/m$^{3}$. Each chain is CG at variable levels of resolution, with a number of CG sites $n_b=1$, $2$, $3$, $4$, $6$, and $10$. The data from the IECG simulations (circles) agree with the theoretical solutions of the excess chemical potential (squares). The horizontal dashed lines are a guide to the eye and show that the excess chemical potential is mostly insensitive to the granularity of the IECG model.}
\label{MCOnb300}
\end{figure}

Figure \ref{MCOnb300} confirms this consistency by demonstrating the independence of the excess chemical potential for IECG chains with $N=300$ monomers across different temperatures and densities: this consistency remains up to a partition of the chain in $n_b=10$ CG sites, where each site represents a $N_b=30$ monomer segment.
The excess chemical potential from the IECG theory and IECG simulations are in remarkable agreement.
Throughout both figures, any notable deviations from the constant trend emerge under the conditions of lowest temperature and highest density, where the insertion of the chain becomes more challenging.

\begin{figure}[tbh]
\includegraphics[width=0.75\columnwidth]{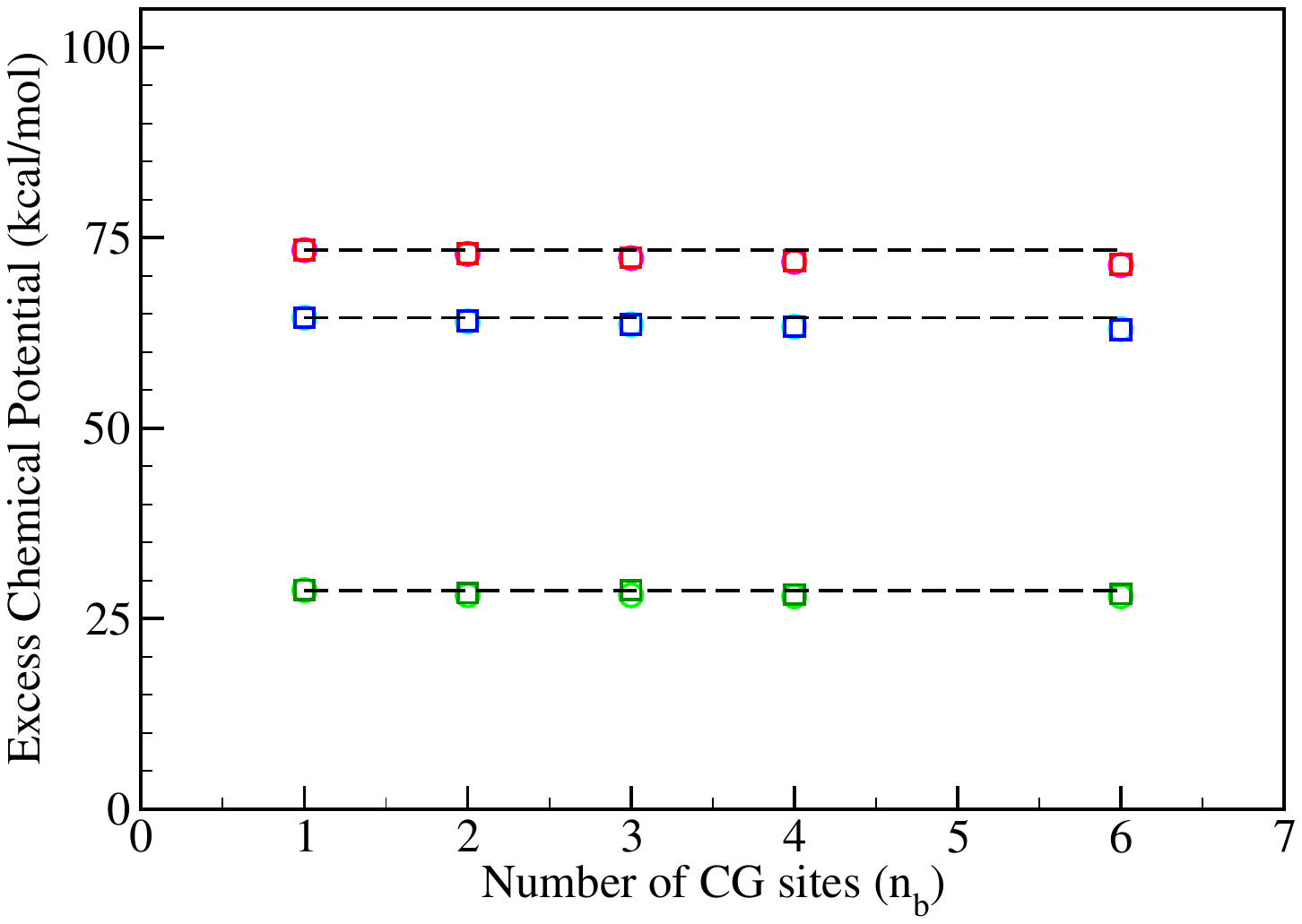}
\caption{Excess chemical potential for a melt of PE chains of degree of polymerization, $N=192$, at variable temperature and density. From the top to the bottom: red symbols $T=513$ K and $\rho=766$ kg/m$^{3}$; blue symbols $T=503$ K and $\rho=766$ kg/m$^{3}$; green symbols $T=503$ K and $\rho=744$ kg/m$^{3}$. Each chain is CG at variable levels of resolution, with a number of CG sites $n_b=1$, $2$, $3$, $4$, and $6$. The data from the MCCCS-CG simulations (circles) agree with the theoretical solution of the excess chemical potential (squares). The horizontal dashed lines are a guide to the eye and show that the excess chemical potential is mostly insensitive to the granularity of the IECG model.}
\label{MCOnb192}
\end{figure}

Figure \ref{MCOnb192} further illustrates the dependence of the excess chemical potential, as determined by the IECG simulations, at various temperatures and densities for the $N=192$ polyethylene chain. This dependence is explored in relation to the number of blobs utilized to partition the chain. Notably, we observe that the excess chemical potential is independent of the CG representation , holding true for a number of CG sites up to $n_b=6$. For this system, each site corresponds to a  segment comprising $N_b=32$ monomers.
Other structural and thermodynamic properties of the IECG melt, as predicted by our Monte Carlo code are reported in the SM.

As anticipated, the thermodynamic consistency inherent to the IECG model showcases robustness, albeit this robustness diminishes as the CG chain is segmented into shorter and shorter segments. Specifically, CG sites that include $N_b=30$ monomers shows some degree of inconsistency. Under such conditions, the distribution of monomers within the chain starts deviating from Gaussian statistics. 
Nonetheless, the consistency of the excess chemical potential remains valid  when the chain is partitioned in a variable number of CG units. This aspect stands as a valuable trait of the IECG model, particularly in the context of multiscale modeling procedures, ensuring that the granularity of the CG model can be flexibly adjusted without introducing undesirable spurious forces.

In the context of a CG polymer modeled as a single soft sphere, where the intramolecular contribution to the excess chemical potential is null, Eqs. \eqref{mu_theory} and \eqref{analytical} are expected to be very accurate. In a multiblob representation, the excess chemical potential will also include an intramolecular contribution, which  incorporates the effect of chain structure change from the ideal to melt state. For a long polymer chain partitioned in a small number of soft blobs, the influence of excluded volume effects is expected to be minimal, so that intramolecular corrections to $\mu^{bb}_{\rm exc}$ are negligible. However, as we progressively divide the polymer into an increasing number of CG sites, each representing shorter chain segments interacting through sharper repulsive blob potentials, this approximation is expected to become progressively less reliable.

\subsection{Testing the code capabilities by increasing the polymer chain length}

\begin{figure}[tbh]
\includegraphics[width=0.75\columnwidth]{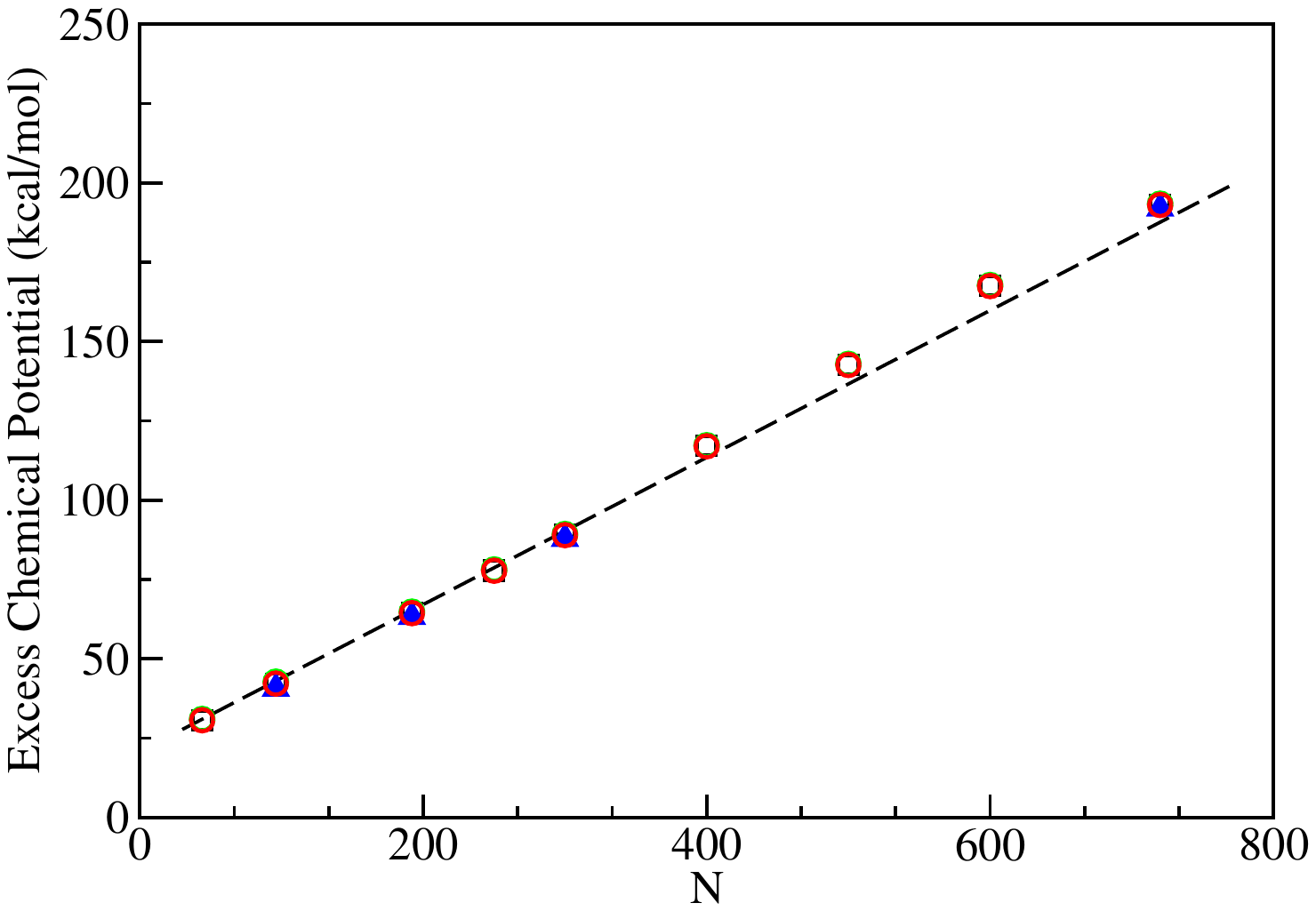}
\caption{Excess chemical potential for a melt of PE in the CG representation as a function of chain length at the density of $766$ kg/m$^3$ and temperature $T=503$ K. Data from MCCCS-CG simulations with $n_b=1$ (red circles), and simulations with $n_b=3$ (solid blue triangles) are shown.  Theory predictions for $n_b=1$ from the numerical solution of the HNC expression, Eq.\eqref{mu_theory} (black squares), and the analytical MSA solution, Eq.\eqref{analytical} (green circles) are also shown, but they are obscured by the simulation data points, being essentially identical in value. The black dashed line is the MSA prediction that uses the fit for $c_0$ in the short-chain limit, extrapolated to long chains, as shown in Figure \ref{C0figure1}.}
\label{MCO1}
\end{figure}

Figure \ref{MCO1} shows the excess chemical potential in both the soft sphere ($n_b=1$) and the three-blob chain ($n_b=3$) representations for a PE melt. As can be seen, altering the degree of CG within the IECG model appears to have no effect, at least within this range of $n_b$. Furthermore, note that $\mu_{exc}^{bb}$ increases with increasing chain length since inserting a longer chain causes a greater increase in the internal energy due to the greater number of, predominantly repulsive at this density,\cite{j_p_hansen_theory_2013} inter-site interactions. The agreement between CG simulation and theory is excellent, with both the HNC and MSA theory predictions being essentially the same. Note that the MCCCS-CG code adeptly computes the excess chemical potential for a PE chain consisting of $N=720$ monomers. This is in stark contrast to conventional insertion methods for atomistic models, which remain constrained to either shorter chains or lower densities.

In the SM we report the excess chemical potential at increasing density and at variable temperature. We observe that an increase in density amplifies the relevance of the repulsive energetic contributions. Consequently, the slope of the excess chemical potential as a function of degree of polymerization becomes more pronounced at higher densities. A similar effect is observed when the simulation temperature is lowered because the insertion of progressively longer chains results in more pronounced chain superposition, accompanied by heightened repulsive interactions.

\subsection{Chemical potential of a single polymer chain in vacuum in the CG representation}

In this section, the computation of the single chain partition function, $Z^{bb}_1$, is further investigated. To that end, it is helpful to examine a pseudo chemical potential, ``$\mu_{\rm intra}$'', which is the average of the Boltzmann factor of the intramolecular potential for a single chain in vacuum:
\begin{equation}
\begin{split}
\label{eqn:muiso}
   \mu_{\rm intra} & =  - k_{\rm B} T \ln{\left[\left< \exp{\left(-\beta U_{\rm intra}\right)} \right>\right]} \\
   & = - k_{\rm B} T \ln{\left[\left(\frac{\int  \exp{\left(-2\beta U_{\rm intra}({\bf r}^{n_b})\right)} d{\bf r}^{n_b} }{\int  \exp{\left(-\beta U_{\rm intra}({\bf r}^{n_b})\right)} d{ \bf r}^{n_b}}\right)\right]},
 \end{split}
\end{equation}
where ${\bf r}^{n_b}$ denotes the configurational space of one polymer chain. In the relative coordinates, the intramolecular energy is given by the sum of three terms
\begin{equation}
\label{eqn:uintra}
\begin{split}
    U_{\rm intra} = &
    \sum_{\alpha}^{n_b-1}U_{\rm bond}(l_{\alpha})      
       +  \sum_{\gamma}^{n_b-2}U_{\rm angle}(\theta_{\gamma})   +  \sum_{\xi\zeta}^{m}U_{\rm eff}^{bb}(r_{\xi\zeta}) ,
    \end{split}
\end{equation}
with $m=(n_b-2)(n_b-3)/2$ as the number of non-bonded interactions. Therefore, in the relative coordinates, Eq.  \eqref{eqn:muiso} can be rewritten as
\begin{equation}
\label{eqn:intra_Gamma}
   \mu_{\rm intra} = -k_{\rm B} T\ln{\left[\frac{\int \exp{\left(-2\beta U_{\rm intra}\right)} d\mathbf{\Gamma}}{\int \exp{\left(-\beta U_{\rm intra}\right)}d\mathbf{\Gamma}}\right]}
\end{equation}
where $\mathbf{\Gamma}$ represents the internal molecular coordinates, $\mathbf{l}$, $\bm{\theta}$, and $\mathbf{r}$. Considering Eqs. \eqref{eqn:muiso} to \eqref{eqn:intra_Gamma} and the Jacobians, one obtains
\begin{equation}
\begin{split}
   \frac{\mu_{\rm intra}}{-k_{\rm B} T} = & \ln{\left[\frac{\int d\mathbf{l}\exp{\left(-2\beta\sum\limits_{i}^{n_b-1}U^{\rm bond}(l_i)\right)}}{\int d\mathbf{l}\exp{\left(-\beta\sum\limits_{i}^{n_b-1}U^{\rm bond}(l_i)\right)}}\right]}  \\ & +
     \ln{\left[\frac{\int d\bm{\theta}\exp{\left(-2\beta\sum\limits_{i}^{n_b-2}U^{\rm angle}(\theta_i)\right)}}{\int d\bm{\theta}\exp{\left(-\beta\sum\limits_{i}^{n_b-2}U^{\rm angle}(\theta_i)\right)}}\right]}  \\ & +
    \ln{\left[\frac{\int d\mathbf{r}\exp{\left(-2\beta\sum\limits_{i}^{m}U^{nb}(r_i)\right)}}{\int d\mathbf{r}\exp{\left(-\beta\sum\limits_{i}^{m}U^{nb}(r_i)\right)}}\right]}
\end{split}
\end{equation}
where each bolded integration measure is an appropriate relative coordinate for the particular intramolecular potential and contains all of the internal coordinates for example, $d\mathbf{l} = l_1^2dl_1l_2^2dl_2 \cdot \dotsc \cdot l_{(n_b-1)}^2dl_{(n_b-1)}$. Each integral in a given term for the CG polymers of this study will contribute equally so we can rewrite each term as a single integral to the appropriate power.
\begin{equation}
\label{muisolate}
  \mu_{\rm intra} = \left( n_b-1 \right)\mu_{\rm bond} + \left( n_b-2 \right) \mu_{\rm angle} + m \mu_{nb},
 \end{equation}
 where
 \begin{eqnarray}
  \mu_{\rm bond} = &&-k_{\rm B} T \ln{\left[\left(\frac{\int  \exp{\left(-2\beta U_{\rm bond}(l)\right)}l^2 dl}{\int  \exp{\left(-\beta U_{\rm bond}(l)\right)}l^2 dl}\right)\right]} \nonumber \\ 
  \mu_{\rm angle} = && -k_{\rm B} T 
     \ln{\left[\left(\frac{\int \exp{\left(-2\beta U_{\rm angle}(\theta)\right) \sin{\theta}\ d\theta}}{\int  \exp{\left(-\beta U_{\rm angle}(\theta)\right)}\sin{\theta}\ d\theta}\right)\right]} \nonumber \\
 \mu_{nb} = && -k_{\rm B} T 
    \ln{\left[\left(\frac{\int  \exp{\left(-2\beta U_{nb}(r)\right)r^2 dr}}{\int  \exp{\left(-\beta U_{nb}(r)\right)}r^2 dr}\right) \right]} 
\end{eqnarray}

In the IECG model, the non-bonded potential is bound and soft. Thus, the change in the intramolecular chemical potential due to non-bonded blob-blob interactions is small relative to the bond and angle contributions. As a consequence, we observe that the intramolecular chemical potential of Eq.(\ref{muisolate}) scales linearly with the number of CG units that make up a chain, $n_b$. 

\begin{figure}[tbh]
\includegraphics[width=0.75\columnwidth]{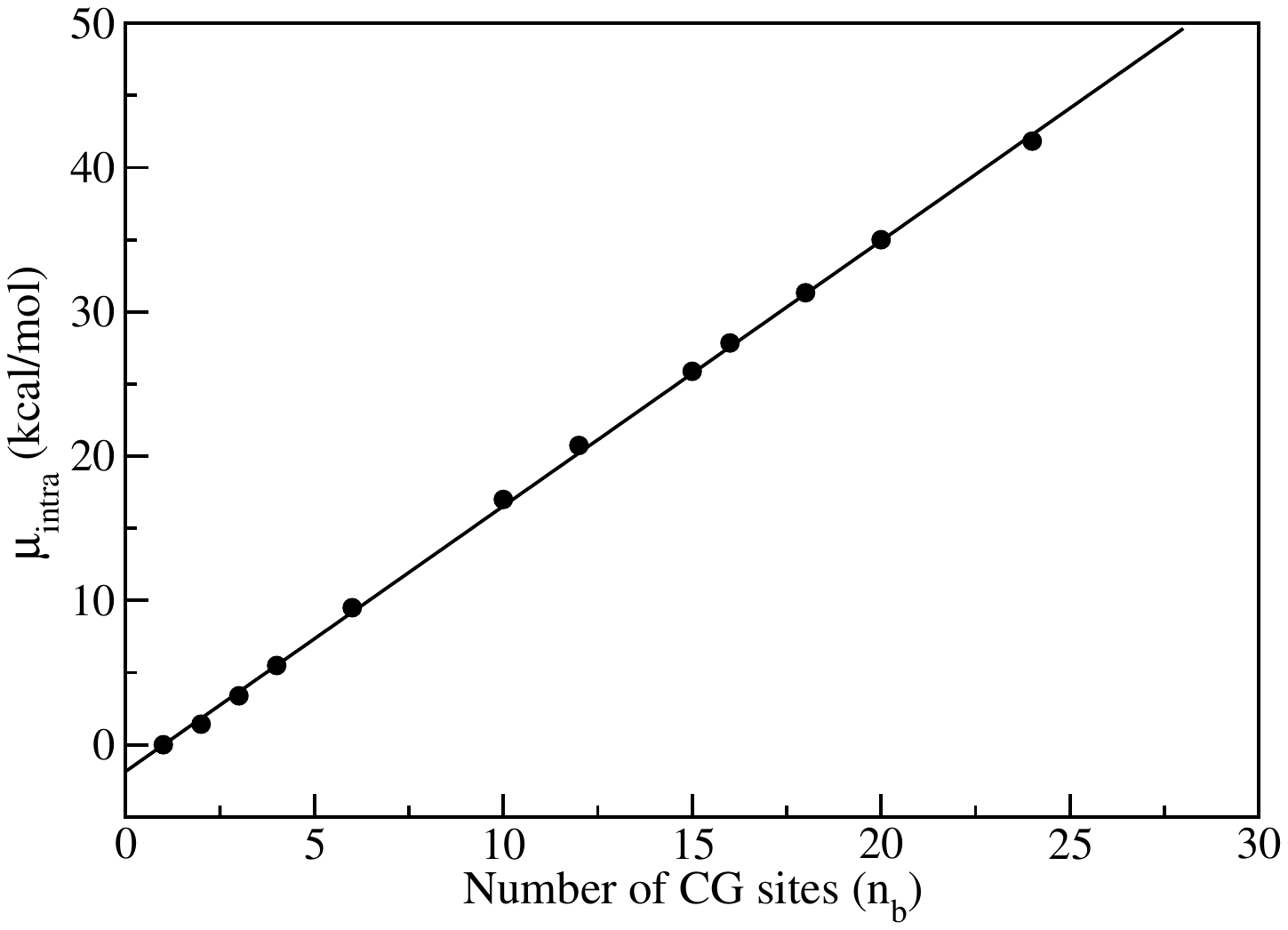}
\caption{The ``intramolecular'' chemical potential estimated from simulations for different levels of CG of an isolated polyethylene chain with degree of polymerization $N = 720$ at a temperature of $T = 503$ K (black circles). The trend-line is a linear fit to the data.}
\label{muintraN720}
\end{figure}

This linear scaling is confirmed by MCCCS-CG simulation data of $\mu_{\rm intra}$ for a CG polymer with degree of polymerization $N = 720$, as shown in Fig. \ref{muintraN720}. 
We calculated $\mu_{\rm intra}$ by simulating a single polymer in vacuum, where the change in internal energy  is measured by insertion of a single chain in the gas phase at the selected temperature. Note that multiple insertions of various single chain conformations were performed to obtain the gas-phase chemical potential of the single polymer chain at a given level of coarse-graining.

The linear scaling was also obtained in calculations of the chemical potential for a polyethylene chain with $N=300$.
Table \ref{tablej} directly compares $\mu_{\rm intra}$ measured in simulations using Eq.\eqref{eqn:muiso} with the theoretical expressions from Eq.\eqref{muisolate} for this particular sample. Each polymer chain is coarse grained by $n_b=1$ up to $10$ CG sites. The agreement between theory and simulations is quite robust, suggesting that the analytical expression of Eq. \eqref{muisolate} gives a faithful representation of an isolated CG polymer chain. 

\begin{table}[h!]
\begin{center}
\caption{Theoretical (Eq.\eqref{muisolate}) and simulated (Eq.\eqref{eqn:muiso}) ``intramolecular'' chemical potential of a single polyethylene chain CG with the IECG approach. Each polymer simulated has a degree of polymerization, $N=300$, and variable number of CG sites, $n_b$. The samples is  at temperature $T= 503$ K in vacuum. The errors for the simulations are around $0.01$ kcal/mol.}
\label{tablej}
\begin{tabular}{ c | c | c }
\hline
\hline
$n_b$ & $\mu_{\rm intra}({\rm sim})$ (kcal/mol)& $\mu_{\rm intra}({\rm theory})$ (kcal/mol)\\
\hline
\hline
$1$  & $0$     & $0$      \\
$2$  & $1.45$  & $1.44$   \\
$3$  & $3.38$  & $3.38$   \\
$4$  & $5.35$  & $5.26$   \\
$6$  & $9.18$  & $9.14$   \\
$10$ & $16.20$ & $16.81$  \\
\hline
\end{tabular}
\end{center}
\end{table}

We note that a similar linear correlations for the chemical potential of the isolated chain at atomistic resolution, where the chemical potential scales linearly with the number of monomers in the polymer chain, was previously derived from simulations of zero-density freely-jointed hard sphere chains\cite{Escobedo1995} and zero density bead-spring chains.\cite{Kumar1991} Thus, the scaling of $\mu_{\rm intra}$ with the number of CG units is consistent with the scaling with its number of monomers as it should, given the definition of the number of CG units $n_b= N/N_b$ with $N_b$ the number of monomers in one chain.

\section{Discussion and Conclusions}
\label{conclusion}
In summary, the Integral Equation theory of Coarse-Graining (IECG) was employed to compute the excess chemical potential of flexible polymer liquids, leading to the derivation of straightforward expressions. In the IECG approach, polymer liquids are modeled as collections of coarse-grained (CG) chains, with each CG site representing a segment of monomers. It is essential that the size of the chain segment exceeds the polymer persistence length to adhere to the Gaussian statistics of monomer pair distributions and meet the theory's prerequisites for direct correlation functions.

A comparison was made between the predictions of the IECG theory for polyethylene melts and data from corresponding simulations that utilized the derived IECG effective potential. This IECG simulation code was a minor extension of the atomistic Gibbs ensemble Monte Carlo software created by Siepmann and co-workers. The study examined conditions involving varying density, temperature, chain length (ranging from $N=44$ to $720$), and the degree of coarse-graining. In all cases, the agreement between theory and simulation regarding the excess chemical potential was found to be excellent. Equally significant, the invariance of the excess chemical potential with respect to the degree of coarse-graining, as predicted by the theory, was confirmed by the CG simulation for practical values of the number of coarse-graining sites.

Developing a procedure for conveniently computing the excess free energy and phase diagrams of dense polymer liquids has long been a goal in polymer physics and engineering. Particle-insertion methods effectively capture the energy change resulting from the insertion of a single molecule and perform optimally at low density and for short chains, where the insertion of a polymer chain is feasible through advanced methods, such as the configuration-bias Monte Carlo method and others. In contrast, field theories focus on the universal behaviors of polymeric systems and are better suited for high-density liquids composed of very long chains, where the specific chemical nature of the polymer is not essential and fluctuations are minimal. There is a region of study between these two extremes, where high-density liquids of long polymer chains are of interest, which needs to be approached by alternative methods. In this region, fluctuations are essential, and differences in the chemical structures of the monomers can result in differences in the chemical potential and phase diagrams. Traditional methods struggle to explore this region effectively. 

Nevertheless, it is possible to compute the excess chemical potential in a simulation by combining particle-based insertion methods with CG models featuring soft potentials. A CG representation in which the entire polymer is depicted as one soft sphere or a collection of soft spheres ensures that the attempted insertion of one polymer is readily accepted, as no sharp repulsive interactions exist during the direct insertion of the soft particle. However, having a CG model with soft interactions is necessary but not sufficient to perform reliable calculations of the excess chemical potential.

A second necessary condition for success is that the CG model accurately predicts the polymer liquid's thermodynamic properties. Two well-established CG models of polymers that represent a chain as a soft sphere are the dissipative particle dynamics (DPD) approach and the IECG model, the latter of which is employed here as mentioned above. Both methods offer computational convenience because the soft sphere representation significantly speeds up CG simulations compared to atomistic simulations. The short-ranged shape of the DPD potential further enhances computational efficiency as it does not require large simulation boxes, unlike IECG. However, the DPD model has been designed to reproduce the dynamics and hydrodynamics of atomistic simulations but does not guarantee consistency in thermodynamic properties, making it unsuitable for calculating polymer phase diagrams.\cite{Frenkel2002}

In contrast, the computational efficiency of the IECG approach primarily arises from the accelerated dynamics predicted by this model, which aligns with atomistic diffusion only once it is  rescaled using the contributions due to the potential of mean force and the solution of the memory function.\cite{Lyubimov2010} The distinct advantage of the IECG model lies in its ability to accurately reproduce both the structure and pressure of the atomistic description, as demonstrated in this work. Consistent pressure is crucial for accurately predicting the excess chemical potential since these two quantities are formally interrelated through the thermodynamics of liquids.

Note that while the analytical solution of the excess chemical potential has been tested for long homopolymer melts in this study, the IECG method has broader applicability. It extends beyond the scope of this particular application because it builds upon the foundation of  atomistic PRISM, which was previously employed to investigate various polymeric systems. The prior extension of the IECG approach to block-copolymers\cite{Sambriski2007b} and polymer mixtures\cite{Yatsenko2004,Yatsenko2005,Dinpajooh2019} opens up the potential for calculating the chemical potential and phase transition for these systems. Thus, the algorithms within the MCCCS-CG code may offer a precise means of predicting the excess chemical potential for a variety of polymeric systems with minimal computational demands.

While it is unfeasible to directly compare the IECG model with atomistic CBMC calculations for these melts of long chains, the IECG model can be directly parameterized using a top-down approach. This involves fine-tuning the IECG model to replicate specific experimental quantities, such as the system's solubility for small molecules dispersed in a polymer matrix. Please note that evaluating the solubility, which is the transfer of a solute between two solvation environments where one is a polymer liquid and the other is a gas phase, requires the use of the IECG approach for mixtures we have previously developed.\cite{Yatsenko2004,Yatsenko2005,Dinpajooh2019}  In this top-down approach, the IECG methodology ensures that important properties remain consistent across various levels of granularity, while significantly outperforming conventional atomistic GEMC methods in terms of speed. These attributes are vital prerequisite for accurately calculating the excess chemical potential for long-chain polymer liquids.

Regarding the potential applications of the IECG method for atomistic chemical potential calculations, there are several feasible avenues. In CBMC algorithms\cite{Siepmann1990} trial configurations are generated with a bias favoring low-energy configurations on a unit-by-unit basis. However, in dense polymer melts of long chains, identifying low-energy configurations becomes more challenging due to the high monomer density, which can result in highly repulsive interactions are short distances when inserting a polymer.
Building on the insights from this study, it is conceivable to enhance existing CBMC algorithms for relatively large polymer chains by integrating CBMC techniques for united atom models with the coarse-grained representation of long-chain polymers. This process involves calculating Rosenbluth weights for growing the united atom units within a given CG site after inserting the CG polymer chain. The objective of growing polymer chains in this way is to enhance the efficiency of CBMC moves, which are currently limited to dense polymer melts with about $30$ monomers, or even fewer depending on the monomer density, using commonly available computational resources. A exciting avenue for reconstructing atomistic chains from CG configurations involves the utilization of machine learning models.\cite{Li2020}
The implementation of such an algorithm remains a subject for future research.

In summary, we have demonstrated that the excess chemical potential remains independent of the level of coarse-graining within the IECG method. This result is contingent on the fact that the excess chemical potential is primarily influenced by the intermolecular component of the internal energy, which remains independent of the chosen level of coarse-graining within the IECG approach. In contrast, a linear dependence on the number of 'blobs' is expected in the ideal part of the chemical potential, which is influenced by intramolecular internal energy. Nevertheless, the intramolecular chemical potential of a long chain in the gas phase, unlike the excess chemical potential in the dense melt,  can be computationally determined at atomistic resolutions when needed.

\section{Supporting Information}
Description of the procedure used to perform Molecular Dynamics simulations at atomistic resolution, employed to validate the coarse-grained model.
Description of the numerical procedure used to calculate the IECG potential.
Validation of the MCCCS-CG simulations by comparison with LAMMPS atomistic simulations, specifically examining the radial distribution function and pressure.
Calculations of the excess chemical potential across various densities and temperatures for polyethylene melts with increasing degree of polymerization, ranging from $N=44$ to $N=300$.

\section{Acknowledgments}
This work was partially supported by the donors of ACS Petroleum Research Fund under New Direction Grant 61648-ND6. M.G.G. served as Principal Investigator on ACS PRF 61648-ND6 that provided support for J.M..
This research was also partially supported by the National Science Foundation Grants No. CHE-1665466 and CHE-2154999.
The computational work was performed on the
supercomputers Comet and Expanse at the San Diego Supercomputer Center, with the support of ACCESS\cite{access} allocation Discover ACCESS CHE100082 (ACCESS is a program supported by the National Science Foundation under Grant No. ACI-1548562).
We thank Dr. M. G. Martin for useful discussions.
M.D. acknowledges the support by U.S. Department of Energy, Office of Science, Basic Energy Sciences, Chemical Sciences, Geosciences, and Biosciences Division, Condensed Phase and Interfacial Molecular Science program, FWP 16249. Pacific Northwest National Laboratory (PNNL) is operated by Battelle for the U.S. DOE under Contract No. DE- AC05-76RL01830.

\bibliographystyle{aipnum4-1}

\end{document}


\title{\centering\Large\bf Supplemental Material to the manuscript entitled: Chemical Potential of a Flexible Polymer Liquid in a Coarse-Grained Representation}

\author{M. Dinpajooh}
\affiliation{Department of Chemistry and Biochemistry, University of Oregon, Eugene, Oregon 97403, USA}
\affiliation{Present address: Physical and Computational Sciences Directorate, Pacific Northwest National Laboratory, WA, 99352 USA}
\author{J. Millis}
\affiliation{Department of Physics, University of Oregon, Eugene, Oregon 97403, USA}
\author{J. P. Donley}
\affiliation{Material Science Institute, University of Oregon, Eugene, Oregon 97403, USA}
\author{M. G. Guenza}
\email{mguenza@uoregon.edu}
\affiliation{Department of Chemistry and Biochemistry, University of Oregon, Eugene, Oregon 97403, USA}
\affiliation{Department of Physics, University of Oregon, Eugene, Oregon 97403, USA}
\affiliation{Material Science Institute, University of Oregon, Eugene, Oregon 97403, USA}
\affiliation{Institute for Fundamental Science, University of Oregon, Eugene, Oregon 97403, USA}
                 
\maketitle

\section{Methods: Molecular Dynamics simulations at atomistic resolution, used to test the coarse-grained description}
\label{methods}

At atomistic resolution, we performed MD simulations in the canonical NVT ensemble with periodic boundary conditions in the three dimensions, using a Nos\'e-Hoover thermostat, and a standard velocity-Verlet integrator. We used the Large-Scale Atomic/Molecular Massively Parallel Simulator (LAMMPS) software package,\cite{Plimpton1995} with the TraPPE-UA forcefield\cite{Martin1998a}, where the hydrogen atoms that are bonded to a carbon are lumped into one united atom. 

\begin{table}[h!]
\caption{United atom TraPPE force field used in Atomistic Molecular Dynamics simulations}
\begin{center}
\label{UApot}
\begin{tabular}{| c | c |c |}
\hline
\multicolumn{3}{|c|}{Bond potential:  \ $U_{bond}=k_b(l-l_0)^2$} \\
\hline
$CH_2-CH_2$ & $k_b kcal \ mol^{-1}$ \text{\AA}$^{-2}$ & $l_0$, \text{\AA} \\
 & 450 & 1.54 \\
\hline
\multicolumn{3}{|c|}{Angle potential:  \ $U_{angle}=k_{\theta}(\theta-\theta_0)^2$} \\
\hline
$CH_2-CH_2-CH_2$ & $k_{\theta} kcal mol^{-1} rad^{-2}$ & $\theta_0$, deg \\
 & 62.1 & 114 \\
\hline
\multicolumn{3}{|c|}{Dihedral potential: $U_{dih}=\sum_{i=1}^3 \frac{C_i}{2}(1+e_i \cos(i\phi))$}\\
\hline
$CH_2-CH_2-CH_2-CH_2$ & $C_i, kcal mol^{-1}$ & $e_i$ \\
 i=1 & 1.4110 & + \\
 i=2 & -0.2708 & - \\
 i=3 & 3.1430 & + \\
 \hline
 \multicolumn{3}{|c|}{Non-bonded potential: $U_{LJ}=4 \epsilon [(\sigma/r)^{12}-(\sigma/r)^{6} ]$}\\
 \hline
 $CH_2$ $---$ $CH_2$ & $\epsilon$, $kcal \ mol^{-1}$ & $\sigma$, \text{\AA} \\
 & 0.0912 & 3.95 \\
 \hline

\end{tabular}
\end{center}
\end{table}

We assumed a Lennard-Jones potential for the interaction between united atoms that belong to different polymers and between united atoms that are more than three atoms apart in the same polymer.  Bonds and angles fluctuate following harmonic potentials. United atoms that are connected by a dihedral angle interact via a OPLS type potential.\cite{Martin1998a} More details on the potential are presented in Table \ref{UApot}. We assumed a cut-off distance of $14$ \text{\AA}, where both the potential and the force were required to go smoothly to zero at the cutoff distance by multiplying the potential by the Mei-Davenport-Fernando (MDF) taper function.\cite{Mei1991a} In all simulations the polymer chains were randomly generated, and overlapping configurations were slowly pushed apart by a soft repulsive potential. After switching on gradually the full non-bonded potential, each system was equilibrated for $200$ ns and then the production runs were performed for about $300$ ns with a timestep of $2$ fs. More details on this procedure can be found in our previous publications.\cite{Dinpajooh2018,Dinpajooh2018a,Guenza2018}

\section{The numerical procedure to calculate the IECG potential}
\label{c0andotherparameters}

The intermolecular IECG potential depends on two parameters, the monomer-monomer direct correlation function at $k \rightarrow 0$, $c_0$, and the polymer mean-square end-to-end distance, $<R^2>$. 

\begin{figure}[htb]
\includegraphics[width=.9\columnwidth]{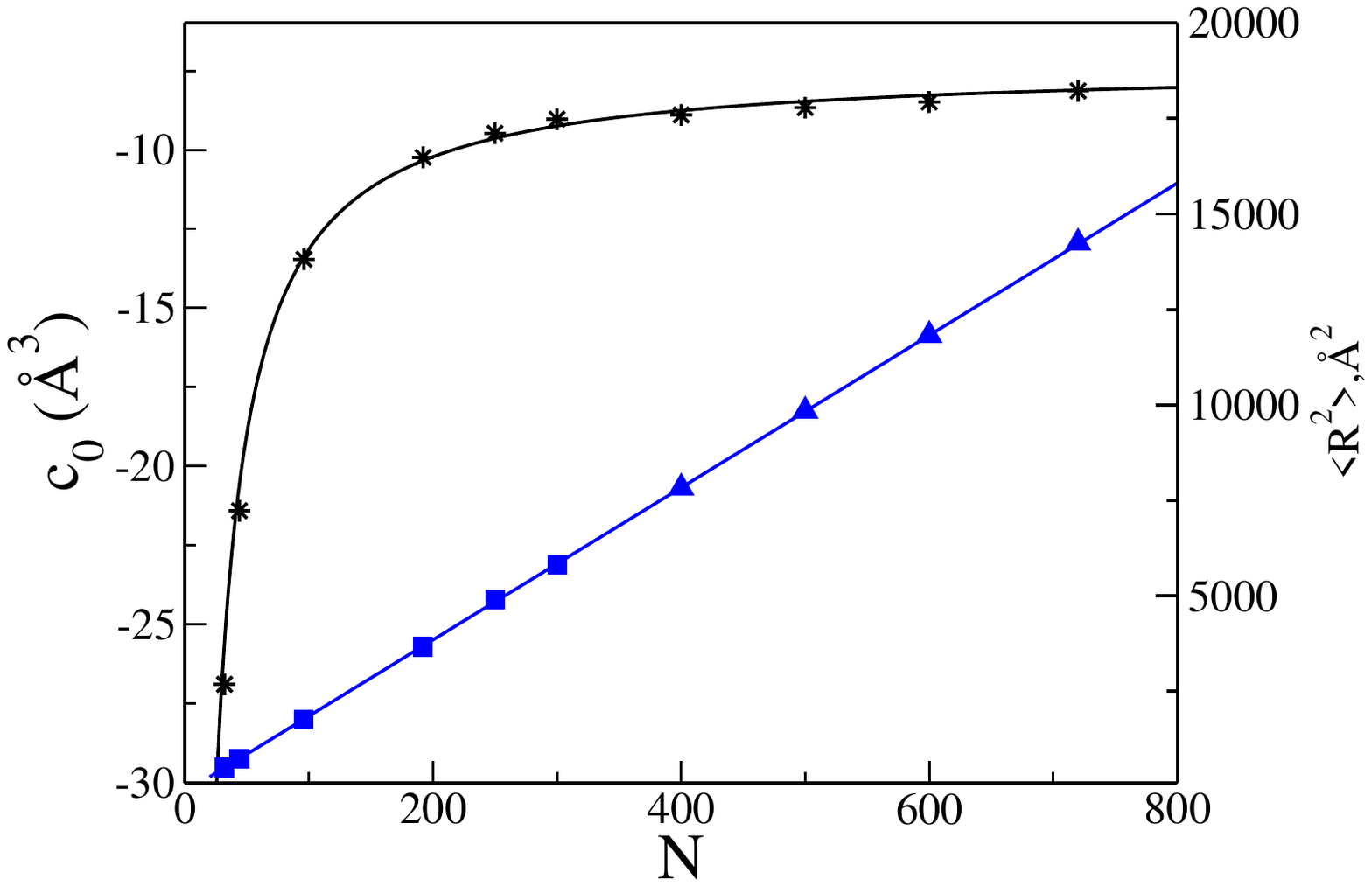}
\caption{Transferability of the IECG method: the values of $c_0$ (black stars) and average square end-to-end distance, $<R^2>$ (blue squares) computed from atomistic simulations of polymer melts of various degree of polymerization $N$,at $T= 503$ K and a density of $\rho=766$ kg/m$^{3}$. The (blue triangles) are values at large $N$ extrapolated from short chain simulations. The end-to-end distance takes a much longer time to equilibrate than the pressure. Data are fitted using the equations $c_0=a + b / N$ and $<R^2> =a' + b' N$. The values for $a$, $b$, $a'$ and $b'$ are $-7.3041$, $-583.45$, $-145.82$, and $19.945$ respectively.}
\label{C0figure1}
\end{figure}

The polymer end-to-end distance enters the definition of the intramolecular distributions, $\hat{\Omega}^{mm}(k)$, $\hat{\Omega}^{bm}(k)$, and $\hat{\Omega}^{bb}(k)$, which enter the equation for the IECG potential,  $U^{bb}_{eff}(r)$,  through $h^{bb}(r)$ and $c^{bb}(r)$. The end-to-end distance is a static property that one can conveniently calculate from a properly equilibrated simulation. For long chains in melts, the end-to-end distance follows the scaling law of unperturbed chains. This is represented in Figure \ref{C0figure1} where $<R^2>$ from computer simulations follows $<R^2> \propto N$. Note that the quality of the potential depends on the smoothness of the intramolecular distributions, which requires coarse-graining of a chain fragment longer than the polymer persistence length. Note also that if the intramolecular pair distributions do not follow Gaussian statistics,  one can still apply the definition of the potential by using as the input parameters to the model the full intramolecular distributions, which can be directly computed from atomistic simulations. 

At fixed thermodynamic conditions, Figure \ref{C0figure1} shows that  the distance follows the statistical properties of polymers in theta conditions, scaling with the degrees of polymerization as $<R^2> =a' + b' N$, as expected. This implies that the end-to-end distance is a parameter that  can be conveniently calculated from atomistic simulations of shorter chains and extrapolated to long chains, with minimal computational requirements. 

The second parameter that enters the equation of the interaction potential is the direct correlation function between monomers, $c_0$. This parameter is calculated from the virial pressure in the high-density approximation, $P= \rho_{ch}k_B T (1-\rho N c_0/2)$. In this case as well, the computational requirements are minimal because the pressure converges rapidly in the atomistic simulations. Thus, to fully define the parameters for the IECG model we need \textit{only a few nanoseconds-long} MD trajectories (less than $3  ns$ for the longest chains simulated here $N=720$). An empirical equation defines the $N$ dependence of $c_0$, as shown in Figure \ref{C0figure1}, which scales  with $N$ as $c_0=a + b/N$. 

The values of $<R^2>$ and $c_0$ follow well-defined equations, where the adjustable parameters ($a$, $a'$, $b$, and $b'$) are defined by fitting the equations with data from atomistic simulations.  Note that because the equations that define the values of the two parameters are known, it is not necessary to perform atomistic simulations for each degree of polymerization of interest, but  only for enough values of $N$ to determine the parameters of the analytical equations with accuracy. This implies that the IECG model can be used to perform simulations for systems that cannot be directly simulated at atomistic resolution. Thus, the IECG potential is transferable and can be used to simulate systems with other degrees of polymerization.

\section{Testing the MCCCS-CG simulations against the LAMMPS atomistic simulations}
\label{checking}

\begin{figure}[htb]
\includegraphics[width=1.15\columnwidth]{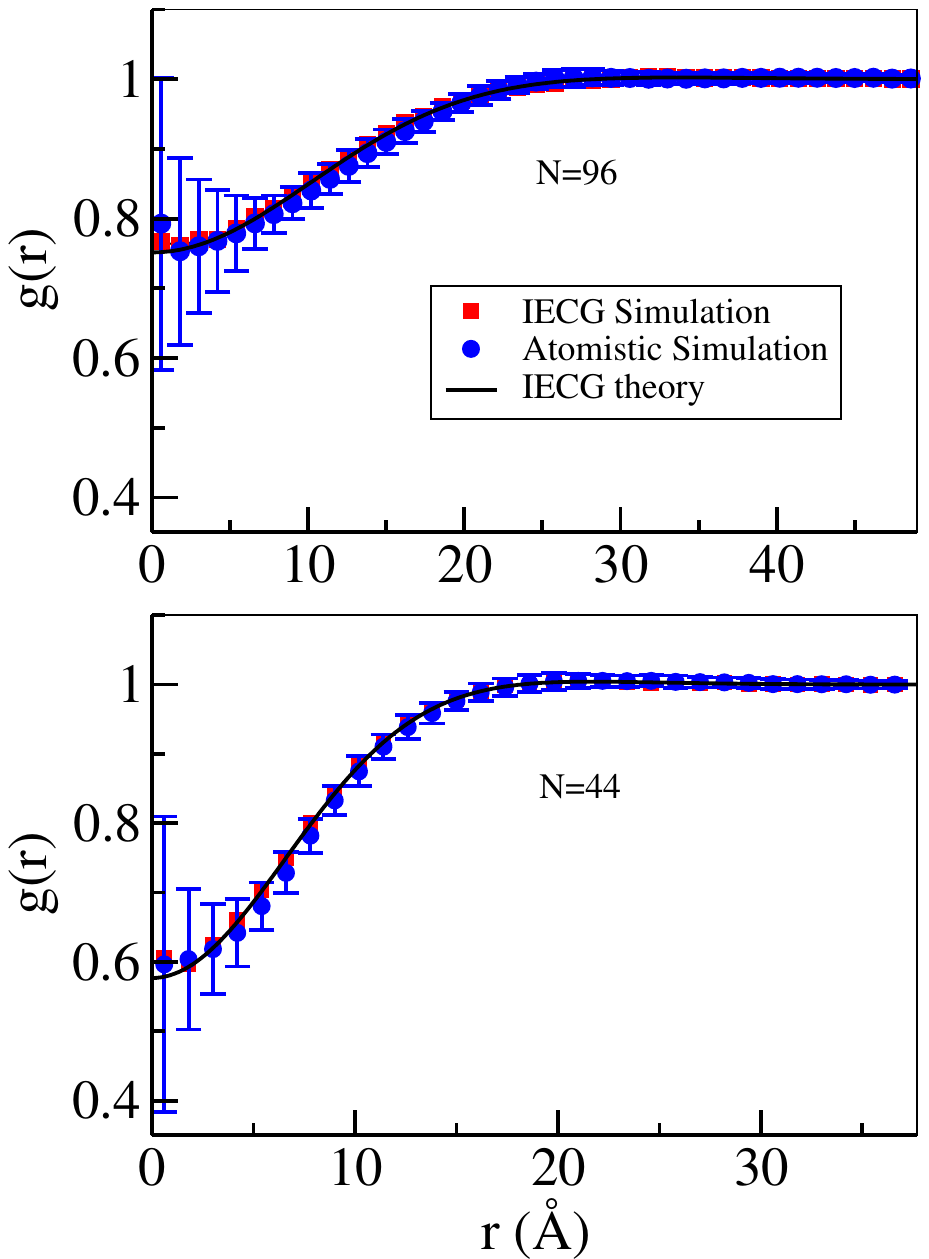}
\caption{Radial distribution functions for PE with N = 44 and N96, T = $503$ K and $\rho = 744$ kg/m$^3$. The black line represents the IECG theory, the red squares are from simulated CG soft spheres and the blue circles represent the atomistic simulation with error bars that were calculated from block averages.}
\label{RDF}
\end{figure}

To test the precision of the MCCCS-CG procedure in constructing and equilibrating the polymeric matrix, while using the IECG representation, we verify that the structure (radial distribution functions) and the equation of state (virial pressure) are predicted to be consistent with the related atomistic simulations. To perform these tests we use the trajectories from the LAMMPS atomistic MD simulations, performed at the identical thermodynamics and molecular conditions (see details in Section \ref{methods}. The consistency observed between the MD and MC simulations ensures that the MCCCS-CG code is numerically correct, and that the equilibration moves that we adopted are sufficient in type and number to create a well-equilibrated polymeric liquid at the given conditions.
As an example, Figure \ref{RDF} shows a comparison between the radial distribution function, $g(r)$, of the equilibrated liquid matrix obtained from the MCCCS-CG code, and the radial distribution function of the corresponding atomistic Molecular Dynamics simulation with the data were mapped onto the coarse-grained representation. The agreement between atomistic and coarse-grained MCCCS-CG structure is quantitative, within the precision of the atomistic simulation data.

\begin{figure}[htb]
\includegraphics[width=1.0\columnwidth]{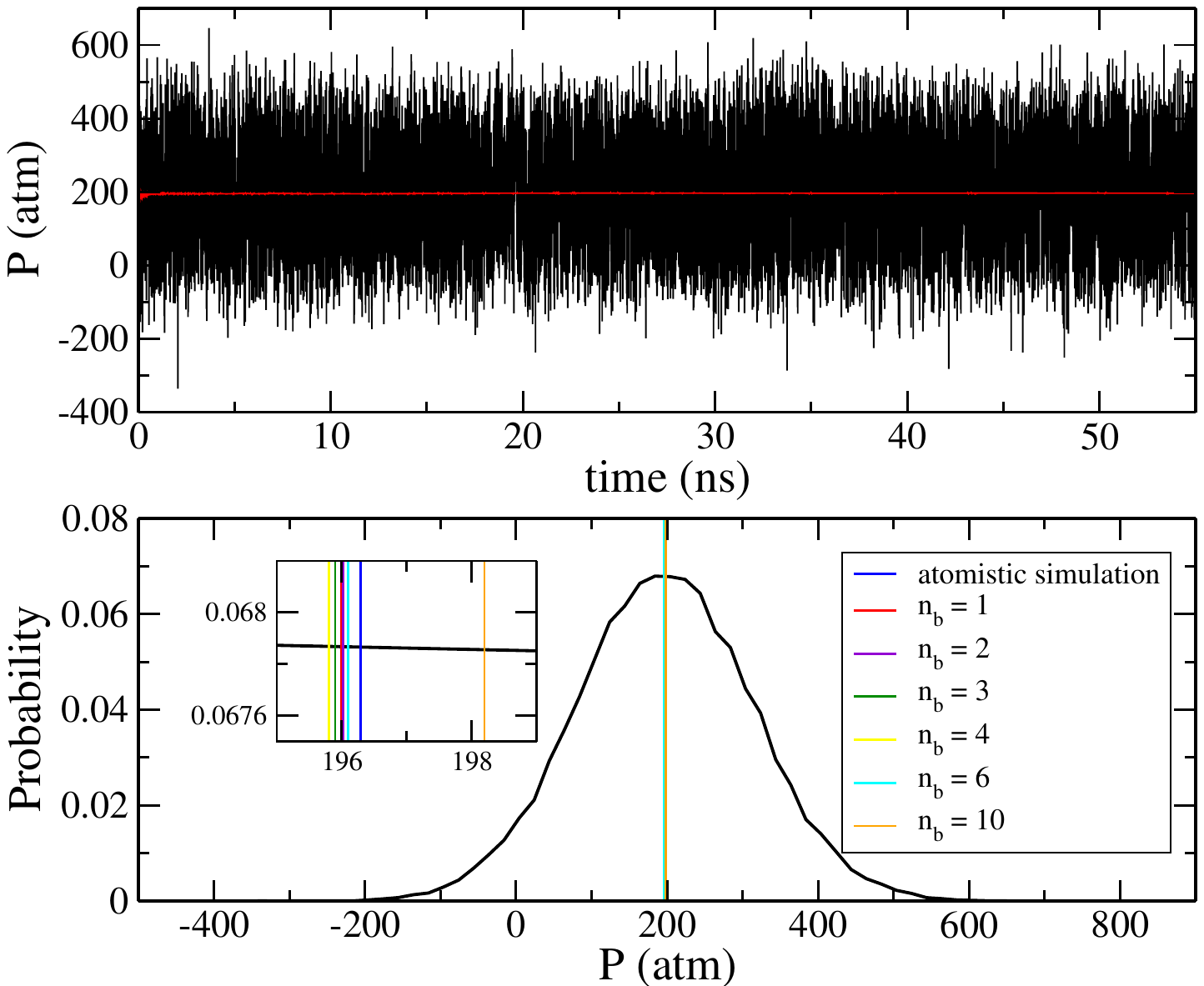}
\caption{Top panel: The instantaneous pressure, in black, and the running average pressure, in red, is shown for an atomistic system of PE with $N = 300$, $T = 473.17$ K and $\rho = 766$ kg/m$^{3}$. Bottom panel: The normalized distribution of the atomistic pressure, the black curve, from the same simulation as the top panel. The average atomistic pressure is shown as a blue vertical line. The average pressure of the same system with different resolutions of CG from the MCCCS-CG code can be seen as vertical red, purple, green, yellow, cyan and orange lines.}
\label{MultiPanelP}
\end{figure}

Figure \ref{MultiPanelP} shows a test of the pressure from the simulated MCCCS-CG IECG system with variable coarse-grained resolution, in comparison with the resulting average pressure from the atomistic MD simulation. The agreement is quantitative, with small differences in the value of the pressure obtained for the coarse-grained model for CG units that contain each more than $30$ PE monomers. The sample with $30$ PE monomers in one blob shows the limitation of the model because, even if the value of the pressure is still within  the numerical error of the atomistic simulation, it starts to diverge from the average atomistic value. This mild disagreement is just an indication that the analytical IECG model, which uses the MSA approximation, needs to be treated with care when the number of monomers in one CG site is small.

While Figures \ref{RDF} and \ref{MultiPanelP} show some examples of these tests, all the data reported in this publication were tested and were found to fulfill these requirements.

The consistency of the IECG virial pressure with the atomistic simulations (see Section \ref{checking}) implies consistency in the physical properties that can be calculated directly from the virial pressure, such as the excess chemical potential. In our previous paper, we presented preliminary calculations of the excess free energy using thermodynamic integration of the pressure.\cite{McCarty2014} In this study, we use the Widom insertion method and Monte Carlo (MC) simulations to directly calculate the excess chemical potential for polymer melts, taking advantage of the IECG description. 

\section{Effect of density and temperature on the excess chemical potential of long polymer chains}

\begin{figure}[htb]
\includegraphics[width=1.\columnwidth]{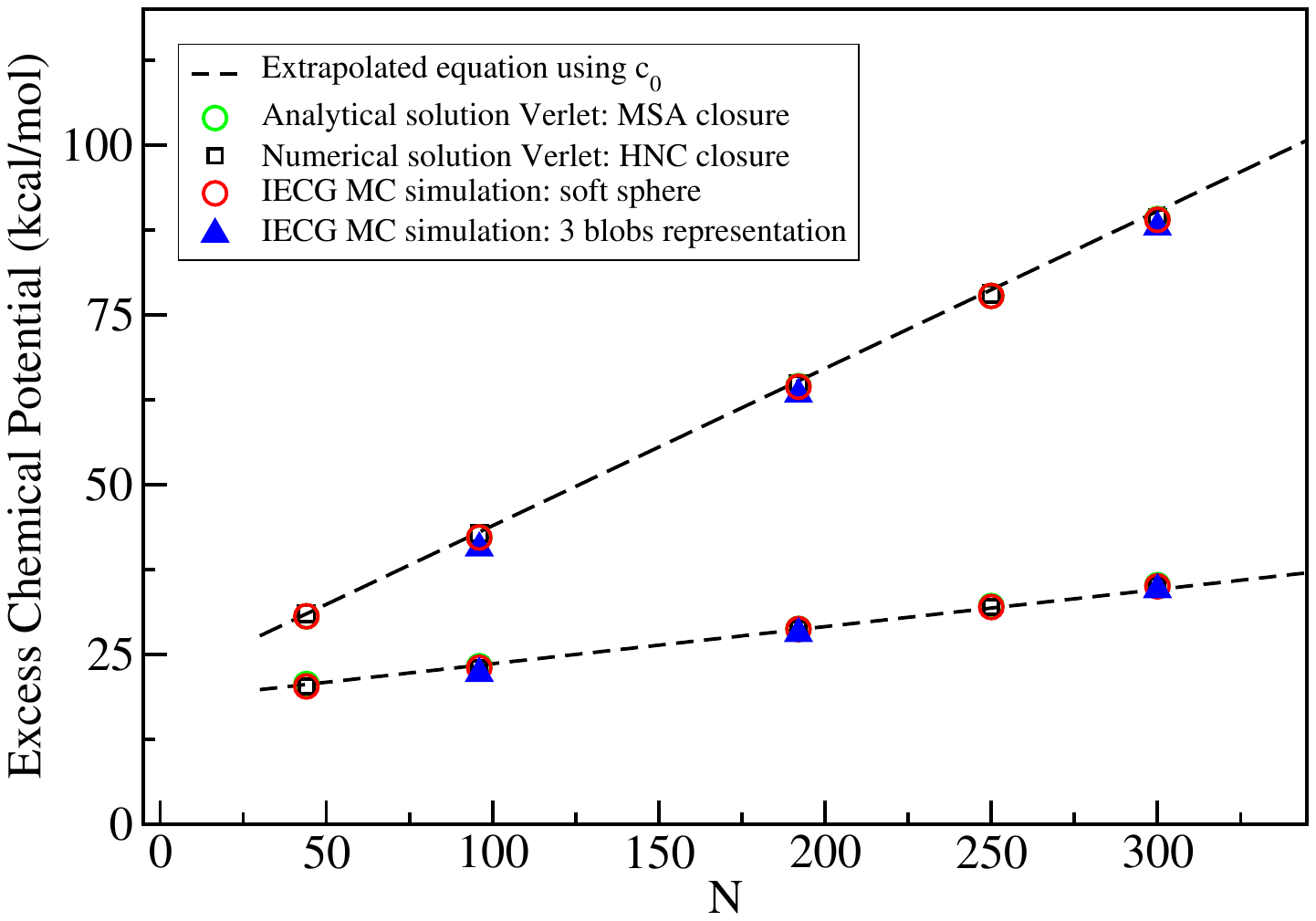}
\caption{Excess chemical potential for a melt of PE in the coarse-grained representation at temperature $T=503$ K and two different monomer densities: $\rho=766$ (top line) and $744$ kg/m$^{3}$ (bottom line). Data from coarse-grained MCCCS-CG simulations with $n_b=1$ (red circles \textcolor{red}{$\boldsymbol\circ$}), and the same with $n_b=3$ (solid blue triangles $\boldsymbol\blacktriangle$) are shown.  Theory predictions for $n_b=1$ from the numerical solution of the HNC expression, (black squares $\boldsymbol\square$), and the analytical MSA solution, (green circles $\boldsymbol\circ$) are also shown, but they are obscured by the simulation data points, being essentially identical in value. The black dashed line (- - -) is the MSA prediction that uses the fit for $c0$ in the short-chain limit, extrapolated to long chains.}
\label{MCOdensity}
\end{figure}

Figure \ref{MCOdensity}  shows the excess chemical potential at increasing degree of polymerization but for two different densities,  $\rho=766$ and $744$ kg/m$^{3}$. As can be seen, an increase in density amplifies the relevance of the repulsive energetic contributions. Consequently, the slope of the excess chemical potential as a function of degree of polymerization becomes more pronounced at higher densities. As for the data in the Main text, the difference between the simulated excess chemical potential for $n_b=1$ and $3$ is of smaller magnitude than the size of the symbols in the figure, which is the inherent error in the data.  

\begin{figure}[htb]
\includegraphics[width=1.\columnwidth]{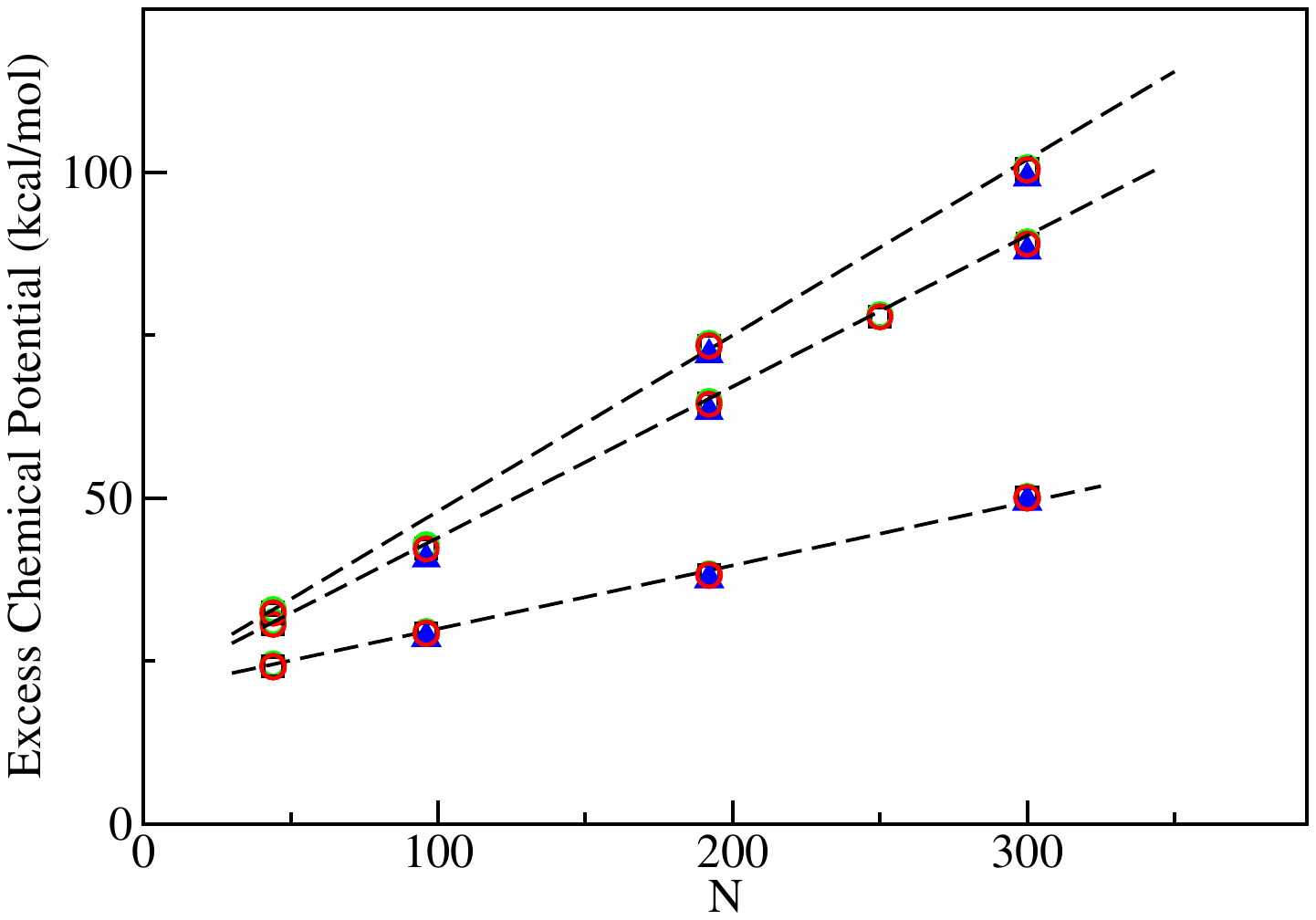}
\caption{Excess chemical potential for a melt of PE at a density of $766$ kg/m$^{3}$ and three different temperatures. From top to bottom, $T=473$, $503$,  and $513$ K. The meaning of the symbols and lines are the same as for \ref{MCOdensity}.}
\label{MCOtemperature}
\end{figure}

Figure \ref{MCOtemperature} shows data that similarly illustrates how the excess chemical potential increases as the simulation temperature is lowered. Notably, the steeper surge in $\mu_{exc}^{bb}$ occurs at the lower temperature, where the insertion of progressively longer chains results in more pronounced chain superposition, accompanied by heightened repulsive interactions.